\documentclass[preprintnumbers, prd, onecolumn, floatfix, superscriptaddress, nofootinbib] {revtex4-2}
\usepackage{setspace}
\usepackage{graphicx}
\usepackage{subfig}
\usepackage{epsfig}
\usepackage{bm}
\usepackage{amssymb}
\usepackage{float}
\usepackage{amsmath}
\usepackage{dcolumn}
\usepackage[colorlinks]{hyperref}
\usepackage[usenames,dvipsnames]{color}
\usepackage{enumitem}
\hypersetup{ breaklinks=true, pdfstartview={FitH},
	colorlinks=true, linkcolor=blue, citecolor=red,
	filecolor=magenta, urlcolor=blue, anchorcolor=green,
	linktocpage=true }

\renewcommand{\(}{\left(}
\renewcommand{\)}{\right)}

\def\doi{http://doi.org}

\def\d{$}
\def\r{\ref}
\def\and{$and$}
\def\g{~}
\def\a{\alpha}
\def\b{\beta}
\def\c{\zeta}
\def\l{\label}

\newcommand{\be}{\begin{equation}}
	\newcommand{\ee}{\end{equation}}
\newcommand{\ban}{\begin{eqnarray*}}
	\newcommand{\ean}{\end{eqnarray*}}
\newcommand{\ba}{\begin{eqnarray}}
	\newcommand{\ea}{\end{eqnarray}}
\newcommand{\no}{\nonumber}

\newcommand{\bc}{\begin{center}}
\newcommand{\ec}{\end{center}}	

\usepackage{xcolor}

\begin{document}
	
	\begin{center}
	
 Observational constraints in late time for an axially symmetric transitioning model with bulk viscous fluid

	\vspace{10mm}
	\normalsize{G. K. Goswami$^1$, Anirudh Pradhan$^2$, Vipin Chandra Dubey$^3$,  }\\
	\vspace{5mm}
    \vspace{2mm}
	\normalsize{$^{1}$Department of Mathematics, Netaji Subhas University of Technology, Delhi, India}\\
	\vspace{5mm}
	\normalsize{$^{2}$Centre for Cosmology, Astrophysics and Space Science (CCASS), GLA University, Mathura-281 406, Uttar Pradesh, India}\\
	\vspace{2mm}
		\normalsize{$^{3}$Department of Mathematics, Institute of Applied Sciences and Humanities, GLA University, Mathura-281 406, 
		Uttar Pradesh, India}\\
	\vspace{2mm}	
		
	$^1$E-mail: gk.goswami9@gmail.com \\
	\vspace{2mm}
	$^2$E-mail:pradhan.anirudh@gmail.com\\
	\vspace{2mm}
	$^3$E-mail: vipindubey476@gmail.com \\
	
	\vspace{10mm}
	
	%\date{}
	%\maketitle
\end{center}

%\begin{abstract}
In this paper, we explore an axially symmetric Bianchi type-I model of the universe with bulk viscous fluid as a source of gravitational field under the framework of Einstein's field equations by assuming barotropic bulk viscous pressure as $-3\zeta H^2$. 
 The model parameters have been estimated with the help of four data sets: The Hubble 46 data set describes Hubble parameter values at various redshifts, Union 2.1 compilation data sets comprise a distance modulus of 580 SNIa supernovae at different redshifts, the Pantheon data set contains Apparent magnitudes of 1048 SNIa supernovae at various redshifts and finally BAO data set of volume averaged distances at 5 redshifts. 
 The observational data is analyzed using the traditional Bayesian methodology, and the posterior distributions of the parameters are obtained using the Markov Chain Monte Carlo (MCMC) technique. To get the best-fit values for the model parameters for MCMC analysis, we use the $ emcee $ package. For parameter estimation, we have also employed the minimizing $\chi^{2}$ function. We also tried to achieve these values statistically using combined data sets from the four described earlier. The OHD+BAO~and~OHD+Pan+BAO+Union combined data sets provide the best fit Hubble parameter value $H_0$ as $66.912 ^{+0.497}_{-0.501})$ Km/s/Mpc and $74.216 ^{+0.150}_{-0.148}$ Km/s/Mpc respectively. We have performed state finder diagnostics to discuss the nature of dark energy. Some other geometrical parameters like the Jerk parameter and the Om diagnostic are also being discussed to clarify the nature of the dark energy model. The study reveals that the model behaves like a quintessence in late time and approaches the $\Lambda$ CDM model.

%\end{abstract}

\smallskip 
{\bf Keywords} : LRS Bianchi-I, $\Lambda$CDM model, Statefinder diagnosis \\
PACS: 98.80.-k \\
	
%%%%%%%%%%%%%%%%%%%%%%%%%%%%%%%%%%%%%%%%%%%%%%%%%% Section 1 %%%%%%%%%%%%%%
\section{Introduction} Two primary features of the universe have been established by current astronomical reflections and modern experimental data from SNe Ia \cite{refN1}$-$\cite{refN6}; CMBR \cite{refN7,refN8}; and WMAP \cite{refN9}$-$\cite{refN12}: (a) the existence of the anisotropic universe at the early stage of evolution, which eventually attains isotropy; and (b) the current universe is expanding along with an accelerating universe. A universe that passes through a transition from past decelerating to current accelerating expansion is indicated by the SNe Ia measurements. Therefore, it is difficult for theorists to adequately justify these observations theoretically. The present world might fit the description of spatially homogenized and isotropic universes provided by Friedmann-Robertson-Walker (FRW) spacetime. However, because of their higher symmetries, these universes offer poor approximations for an early universe and do not accurately describe the matter of the early universe. These models, which exhibit an anisotropic nature in early periods and approaches to isotropy in later times, are thus more applications for the general framework of the universe's evolution. Bianchi-type or an Axially Symmetric space-time provides a good foundation for this. \\

Axially symmetric spacetime refers to a type of spacetime geometry that has rotational symmetry around a central axis, meaning that its properties remain unchanged under rotations about this axis. In studying transitional cosmological models—models that describe the universe's evolution across different phases (e.g., from an inflationary phase to a matter-dominated phase)—axial symmetry can serve as a simplifying assumption. Transitional models often involve phases with different symmetry properties, and axial symmetry can capture important details of these transitions without requiring full isotropy (symmetry in all directions). Herrera et al. \cite{refN13} explore axially symmetric spacetimes and the physical interpretation of sources that generate them, emphasizing their importance in general relativity. The comprehensive book \cite{refN14} discusses exact solutions, including axially symmetric spacetimes, providing detailed insights into their properties and applications in various cosmological models. Coley et al. \cite{refN15} examined cosmological models with symmetry properties, including axially symmetric spacetimes, and discussed their role in transitional cosmology and structure formation. Carmeli et al. \cite{refN16}  specifically address axially symmetric cosmological models and their significance in describing anisotropic properties during different phases of cosmic evolution. The review paper by Joshi et al. \cite{refN17} includes discussions on axially symmetric spacetimes, particularly about gravitational collapse, and their impact on understanding spacetime singularities. These references provide a foundation for understanding axially symmetric spacetimes and their importance in transitional and anisotropic cosmological models, making them valuable resources for both theoretical and observational cosmology. \\

It has long been known that viscosity coefficients are important in cosmological modelling. Every time a system loses its thermal equilibrium condition, an effective pressure is created to restore its thermal stability, according to hydrodynamics theory. A manifestation of such an effective pressure can be seen in the bulk viscosity of the universe's matter content. In their cosmological models, Nojiri and Odintsov \cite{refN18,refN19} considered dark energy (DE) and dark matter (DM) as imperfect fluids. One specific example of the concepts discussed in \cite{refN18,refN19} is viscous fluids. Viscous pressure associated with bulk viscosity can act like a negative pressure in a cosmological model. Negative pressure is a defining feature of dark energy since it opposes gravitational attraction and leads to cosmic acceleration. A few studies in recent years have examined the hypothesis that a type of viscous fluid is responsible for the late-time acceleration \cite{refN20,refN21,refN22}. The theory of evolution of the viscous pressure under the context of late-time acceleration of the cosmos has been used to analyse a universe full of bulk viscous matter in these references \cite{refN20,refN21,refN22}.  It should be noted at this point that the late-time accelerated phase is not just the universe's accelerated phase. In the early stages of the cosmos, there was another acceleration phase known as an inflationary scenario \cite{refN22}. Dissipative effects, such as bulk and shear viscosity, are believed to be important at this early stage of evolution \cite{refN23}. In their work, Chimento et al. \cite{refN24} claimed that a combination of a quintessence substance and a cosmic fluid with bulk-dissipative pressure might lead to an accelerated expansion of the universe. Additionally, \cite{refN24} reported that the aforementioned process entails a series of dissipative processes.  The first attempts to develop the theory of relativistic dissipative fluids were documented in the works of Eckart \cite{refN25} and Landau and Lipshitz \cite{refN26}. Under the influence of bulk viscosity and potential future singularities, Brevik et al. \cite{refN27} examined the unification of inflationary and late-time acceleration. In their proposed method, According to the Eckart theory of bulk viscosity, Silva et al. \cite{refN28} showed that an extended $\Lambda$-CDM model includes a flat universe with a dissipative non-extensive viscous dark matter component.\cite{refN29,refN30,refN31,refN32,refN33,refN34,refN35,refN36,refN37,refN38,refN39,refN40} are notable recent works on viscous cosmology. Some studies \cite{refN31,refN32,refN35,refN37,refN41} have shown that bulk viscous models can be compatible with current cosmological observations, including supernova data and cosmic microwave background measurements, though more precise observations are needed to confirm whether such models provide a complete alternative. Using a new $f(R, L_{m}, T)$ gravity model, Sharma et al. \cite{refR1} investigate the universe's late-time acceleration phase, obtain the corresponding analytical solution, and then constrain arbitrary parameters of the solution by taking into account the Panthoen+SH0ES data and cosmic chronometers. They then examine the behavior of the obtained constrained solution using the deceleration, effective equation of state, and Om diagnostic test. Extending the conventional $\Lambda$CDM model to include dissipative effects in a causal viscous framework and finding an analytical solution for the Hubble parameter is a difficult task in the literature. The authors \cite{refR2} provide a unique solution for the Hubble parameter by presenting a new form for the bulk viscous coefficient associated with dark matter. The authors \cite{refR3} concentrated on teleparallel gravity, where the fundamental geometric quantity representing gravitational interactions is torsion, and studied the impact of altering bulk viscosity coefficients. Here the $f (T)$ model, which includes bulk viscous matter, has the ability to explain cosmic acceleration, according to their findings. The authors \cite{refR4} study the impact of bulk viscosity on late-time cosmic acceleration in an extended $f (Q, L_{m})$ gravity framework. The non-metricity $Q$ is not minimally coupled with the matter Lagrangian $L_m$.  The author argues that bulk viscosity plays a crucial role in explaining faster expansion in alternative gravity theories. Koussour et al. \cite{refR5} study the function of bulk viscosity in $f(Q,T)$ gravity in understanding late-time cosmic acceleration. To evaluate the model, the authors calculate its exact solution and use Hubble parameter $H(z)$ and Pantheon + SNe Ia data to estimate parameters.
\\

Our study aims to examine the effects of the bulk viscosity coefficient $\xi$ in the standard cosmic pressure on the universe's evolution phase. Barotropic bulk viscous pressure is assumed to be $-3\zeta H^2$. This produces negative pressure, which leads to the modeling of an accelerating universe after it has previously gone through a decelerating phase. The assumption of barotropic bulk viscous pressure of the form $-3\zeta H^2$
is indeed common in cosmology and astrophysics, particularly when dealing with models of the universe's expansion, such as those found in cosmological fluid dynamics or in modified gravity theories. However, the deeper physical motivation for this specific choice is often not fully discussed. Let's break this down a bit.
\begin{itemize}
	\item \textbf{Barotropic Viscous Pressure:}  A barotropic fluid is one where the pressure p is a function of the energy density $\rho$ alone, i.e., p=p($\rho$). The bulk viscosity, $\xi$, is typically associated with a dissipation mechanism that resists the expansion or contraction of a fluid due to internal friction. In a cosmological context, it describes the resistance to the expansion of the universe or any other fluid-like entity in space-time.\\
	The form of the bulk viscous pressure $-3\zeta H^2$ relates the viscous pressure to the Hubble parameter H, where $H=\frac{\dot{a}}{a}$  is the rate of expansion of the universe.
    The choice of the bulk viscous pressure as  $-3\zeta H^2$
	can be understood in the following ways as a physical motivations:
	\item \textbf{Thermodynamic Consistency:}  The bulk viscous pressure generally arises in the context of the energy-momentum tensor. For a viscous fluid in an expanding space, the form
	 $-3\zeta H^2$ comes from the evolution equations of the energy density and pressure in a cosmological setting. It can be derived from the conservation of energy and momentum and reflects how dissipation in the fluid affects the expansion.
	 \item \textbf{Hydrodynamical Models:} In the context of a perfect fluid with viscosity, the viscous pressure can be related to the rate of expansion of the fluid. In an expanding universe, the pressure due to bulk viscosity typically behaves like( $\sim -3\zeta H^2$) where the -3 factor comes from the spatial geometry of the universe (in a FRW metric) and from the standard energy-momentum tensor for viscous fluids in cosmology. This factor of 3 ensures that the form of the pressure is consistent with the isotropy and homogeneity of the universe on large scales.
	 \item\textbf{ Dimensional Considerations:}  The form $ -3\zeta H^2$ has the right dimensions for a pressure term in the context of cosmology. Since  H has dimensions of inverse time, $H^2$ has dimensions of time$^{-2}$, and multiplying it by a viscosity term 
	$\zeta$ with dimensions of  time$^{3}$ gives the correct dimensions for pressure ( mass/
	 length$^2$).
	\end{itemize}
 The structure of the paper is as follows: In section I, the current cosmological situation is discussed.  The Einstein field equations for an axially symmetric Bianchi type-I with a bulk viscous fluid as the gravitational field source are presented in section 2. The field equations and the Hubble, deceleration, and Om parameters have also been solved in this section. In section 3, the model is made under the constraints of the four data sets. The Hubble 46 data set(OHD) describes Hubble parameter values at various redshifts (\cite{refN41}). Union 2.1 compilation data sets(Union) comprising of distance modulus of 580 SNIa supernovae at different redshifts (\cite{refN42}). The Pantheon data set(Pan) which contains Apparent magnitudes of 1048 SNIa supernovae at various redshifts (\cite{refN43}) and finally BAO data set(BAO) of volume averaged distances at 5 redshifts (\cite{refN51}). These data sets and their combinations were used to estimate the model parameters $H_0$, $l$ and $\zeta.$ The OHD+BAO~and~OHD+Pan+BAO+Union combined data sets provide the best fit Hubble parameter value $H_0$ as $66.912 ^{+0.497}_{-0.501}$ Km/s/Mpc and $74.216 ^{+0.150}_{-0.148}$ Km/s/Mpc respectively. The model's various geometrical and physical properties were also investigated in sections 4 to 9. In particular, we have obtained the present age of the universe, its present density, and the transitional redshift where the universe entered the accelerating phase after passing through the deceleration in the past. We have performed state finder diagnostics to discuss the nature of dark energy. Some other geometrical parameters like the Jerk parameter and the Om diagnostic are also being discussed to clarify the nature of the dark energy model. The study reveals that the model behaves like a quintessence in late time and approaches the $\Lambda$ CDM model. Our developed model is found to be in good agreement with observational findings. The last section contains concluding remarks and highlights of the paper.

%%%%%%%%%%%%%%%%%%%%%%%%%%%%%%%%%%%%%%%%%% Section 2 %%%%%%%%%%%%%%%%%%%%%%%%%%%%%%%%%%%%%
\section{  Metric and the Field Equations}
We begin with the following LRS Bianchi type I metric given as:
\begin{equation}\label{1}
	ds^2= -c^2 dt^2 + \alpha^2 (dx)^2 + \beta^2 \left( dy^2+ dz^2 \right), 
\end{equation} 
where $c$ is the speed of light and the scale factors `$\alpha$ and $\beta$' are time-dependent in a spatially anisotropic universe with homogeneous expansion.  The Einstein field equations connect the distribution of matter and energy through the energy-momentum tensor $T_{ij}$ to the geometry of spacetime which is described by the following equation:

\begin{equation}\label{2}
		R_{ij}-\frac{1}{2} g_{ij} R = \frac{8\pi G}{c^4}T_{ij}
\end{equation}
We consider the following energy-momentum tensor as that of fluid having bulk viscosity given as: 
\begin{equation}\label{3}
	T_{ij} = (p_{eff}+\rho)u_{i}u_{j}+p_{eff} g_{ij},
\end{equation}
where, the effective pressure $p_{eff} =p-3\zeta H^2$ consists of  proper pressure $p$ and  barotropic bulk viscous pressure $3\zeta H^2$.  Here, $H$ is the Hubble parameter representing a rate of expansion of the universe.,  $\zeta$ is the coefficient of bulk viscosity and $ \rho $ is the energy density of the viscous fluid. At present, matter is dust-dominated, so we take proper pressure $p=0.$ 
The following are the Einstein field equations for the LRS Bianchi type-I universe:
\begin{equation}\label{4}
	2 \frac{\ddot{\beta}}{\beta}+ \frac{\dot{\beta}^2}{\beta^2}=-\frac{8\pi G}{c^2}(p-3\zeta H^2)
\end{equation}
\begin{equation}\label{5}
	\frac{\ddot{\alpha}}{\alpha}+\frac{\ddot{\beta}}{\beta}+ \frac{\dot{\alpha} \dot{\beta} }{\alpha \beta}=-\frac{8\pi G}{ c^2}(p-3\zeta H^2)
\end{equation}
\begin{equation}\label{6}
	2\frac{\dot{\alpha} \dot{\beta}}{\alpha \beta}+\frac{\dot{\beta}^2}{\beta^2}=\frac{8\pi G}{ c^2} \rho,
\end{equation}
where dot and double dots are the first and second-order derivatives with respect to the domain i.e. time. The  volume for the model is given by  $ V =\alpha \beta^{2} $, the  scale factor is considered as $ a =\left(\alpha \beta^{2}\right)^{1/3} $ and accordingly, the Hubble's parameter is defined  as
\be \l{6a}
 H = \frac{1}{3} \left(\frac{\dot{\alpha}}{\alpha}+2\frac{\dot{\beta}}{\beta}\right)= \frac{\dot{a}}{a}.
 \ee
 Now we solve the field equations Eqs. (\ref{4}) $-$ (\ref{6}) in the following manner.\\

From Eqs. (\ref{4}) and (\ref{5}), we find
\begin{equation}\label{7}
	\frac{d}{dt}\left(\frac{\dot{\alpha}}{\alpha}-\frac{\dot{\beta}}{\beta}\right)+\left(\frac{\dot{\alpha}}{\alpha}-\frac{\dot{\beta}}{\beta}\right) \left(\frac{\dot{\alpha}}{\alpha}+
	2\frac{\dot{\beta}}{\beta}\right)=0
\end{equation}
On integrating it, we get
\begin{equation}\label{8}
	\frac{\dot{\alpha}}{\alpha}-\frac{\dot{\beta}}{\beta} =\frac{c_1}{a^3}
\end{equation}

Further, on solving Eqs. (\ref{6a}) and (\ref{8}), we get
\begin{equation}\label{9}
	\frac{\dot{\alpha}}{\alpha}=\frac{\dot{a}}{a}+\frac{2}{3} \frac{c_1}{a^3}
\end{equation}
\begin{equation}\label{10}
	\frac{\dot{\beta}}{\beta}=\frac{\dot{a}}{a}-\frac{1}{3} \frac{c_1}{a^3}
\end{equation}
The field Eqs. (\ref{4}), (\ref{5}) and (\ref{6}) are simplified as follows, on using  Eqs. (\ref{9}) and  (\ref{10})
\begin{equation}\label{10a}
\frac{2 \ddot{a}(t)}{a(t)}+\frac{\dot{a}(t)^2}{a(t)^2}+\frac{{c_1}^2}{3 a(t)^6} = -\frac{8\pi G}{ c^2}(p-3\zeta \frac{\dot{a}(t)^2}{a(t)^2})
\end{equation}
and
\begin{equation}\label{10b}
\frac{3 \dot{a}(t)^2}{a(t)^2}-\frac{{c_1}^2}{3 a(t)^6}  = \frac{8\pi G}{ c^2} \rho
\end{equation}
To find the values of unknown model parameters $H_0$, $\zeta$, and $c_1$, we shall estimate them with the help of various data sets which will be described in the coming subsections. For this, we will need to express and solve the Einstein field equations (\ref{10a}) and  (\ref{10b}) in terms of redshift z instead of time. We use the following transformation equations between scale factor a(t) and z.
\begin{equation*}
\frac{a_0}{a(t)} =1+z, \dot{H}(t) = -(1+z) H(z)   \dot{H}(z)
\end{equation*}
On using these, we get the replacement of Eq. (\ref{10a}) as
\begin{equation*}
 \frac{\text{c1}^2 (z+1)^6}{3a_0^6} - 2 (z+1) H(z) \dot{H}(z)+3 H(z)^2=3 \frac{8\pi G \xi}{ c^2}  H(z)^2 .
\end{equation*}
To solve this, we obtain
\begin{equation}\label{11}
	H(z)=\frac{{H_0} (z+1)^{3/2} (z+1)^{-\frac{3 \zeta }{2}} \sqrt{{l}^2 \left((z+1)^{3 \zeta +3}-1\right)+9 (\zeta +1)}}{3 \sqrt{\zeta +1}}, 
\end{equation}
where $l = \frac{c_1}{H_0a_0^3},\zeta=\frac{8\pi G \xi}{ c^2}$ and $H_0$ denotes the present value of the Hubble parameter.
The deceleration parameter (DP) for the model can be expressed as:

\begin{equation}\label{12}
	q= -1 + \frac{(z+1)\dot{H}(z)}{H(z)}
\end{equation}
The equation of state parameter(EOS) is given as:
\begin{equation}\label{13}
	\omega (z) = \frac{p_{eff}}{\rho} = -1 + \frac{(z+1) \dot{H}(z)}{3 H(z)}
\end{equation}
The Hubble parameter $`H'$ and deceleration parameter $`q'$ are the important physical quantities for describing the universe's evolution.
$H$ plays a vital role in explaining the universe's expansion and is also very useful in estimating its age. 
On the other hand, the deceleration parameter describes the phase transition (acceleration or deceleration) during the universe's evolution.

%%%%%%%%%%%%%%%%%%%%%%%%%%%%%%%%%%%%%%%%%%%%%%%%%%%%%%%%%%%%%%%%%%%%%%

\section{Estimations of model parameters $H_{0}$, l, and $\zeta$ from available SNIa, Hubble, and BAO datasets.}
In this section, we have estimated model parameters by using the four datasets, namely, (a) Estimations of Model parameters from Hubble 46 dataset (b) Estimations of Model parameters from Supernova SNIa Union 2.1 Compilation 680 data set (c) Estimations of Model parameters from Supernova SNIa 1048 Pantheon data set and (d) Baryon Acoustic Oscillations (BAO) Analysis. The Hubble 46 dataset comprises distance and redshift measurements for galaxies. By calculating model parameters from this data, we can improve our understanding of the Hubble constant ($H_{0}$), which describes the pace at which the universe is expanding. The estimated parameters can be compared with values obtained from other datasets, such as Cosmic Microwave Background (CMB) data from Planck or Baryon Acoustic Oscillation (BAO) measurements, helping to resolve tensions in cosmological measurements. The Supernova SNIa Union 2.1 Compilation (680 data points) is frequently used in cosmology because it contains a large, high-quality dataset of Type Ia supernovae (SNe Ia), which are used as standard candles to measure cosmic distance. The Pantheon dataset contains 1048 SNe Ia over a wide redshift range ($0.01 < z < 2.3$). The Pantheon dataset (1048 SNe Ia) is a gold standard for evaluating model parameters in cosmology. BAO measurements provide precise estimates of the Hubble parameter $H(z)$ at different redshifts. Surveys like BOSS, eBOSS, SDSS, DESI, and WiggleZ offer accurate BAO measurements across a wide range of redshifts ($0 < z < 3$), enhancing cosmic constraints.
\\
\subsection{Estimations of Model parameters from Hubble 46
	dataset.}\l{A}
We consider the  Hubble parameter table\cite{refN41} consisting of
46 data set of of observed values of \g \d H \d\g for various
redshift in the range \g \d 0 \leq z \leq 2.36\d\g with
possible error in observations. We use this data set to estimate the model parameters \g $H_{0}$, l and $\zeta$  \g in multiple ways. First, we fit
Hubble parameter function given by Eq.(\r{11}) to these data
by a method of least squire and estimates model parameters. There
after assuming these estimated values as initial guesses, we use
method of least $\chi^{2}$ to make more refined estimations.
Finally, we carry Monte-Carlo simulations MCMC method to further
refine the estimations by assuming $\chi^{2}$ estimations as
initial guess. We recall that $\chi^{2}$ formula for the Hubble-function is as follows:
\be
\chi^{2}(H_0, l, \zeta) =
\sum\limits_{i=1}^{46}\frac{(H_{th}(z_{i},H_0, l, \zeta) -
	H_{ob}(z_{i}))^{2}}{\sigma {(z_{i})}^{2}}.
\l{chiHub}\ee
Our estimations are shown in the following Table 1
\begin{table}[H]\l{table1}
	\centering 
	{\begin{tabular}{c|c|c|c|}
			\hline\hline
			Parameters & Least square & Least\g\d \chi^2\d\g & MCMC
			simulation\\
			& Estimations & Estimations & Estimations\\
			\\
			\hline
			\\
                \d H_{0}\d & $60$  & $66.83\pm 0.509$ & $66.836 \pm0.505$
			\\
			\\
                \hline
			\\
			    l  & $0.0459$ & $0.239\pm 0.024$ & $0.239 \pm 0.024$\\
			\\
			\hline
			\\
			\d\zeta\d & $0.274$ & $0.523\pm0.008$ & $0.523 \pm0.008 $\\
			\\
			\hline\hline
	\end{tabular}}
	\caption{\small{ The best-fit values of the model parameters
			 $H_{0}$, l and $\zeta$ for our model $H(z)$ Hubble
			curve constraint with the 46  observed Hubble data set.}}
\end{table}

We consider the  Hubble parameter table consisting of 46 data set of observed values of ~$ H $~ for various
redshift in the range~$ 0 \leq z \leq 2.36~$ with
possible error in observations. We use this data set to estimate the model parameters ~ $H_{0}$, l and $\zeta$  ~in multiple ways. First, we fit
Hubble parameter function given by Eq. (\r{11}) to these data
by a method of least squire and estimates model parameters. There
after assuming these estimated values as initial guesses, we use
method of least $\chi^{2}$ to make more refined estimations.
Finally, we carry MCMC method to further
refine the estimations by assuming $\chi^{2}$ estimations as
initial guess.\\

In all other data sets, we have taken MCMC priors on same ground which means that the parameter values obtained  on the basis of minimum $\chi^2$ are treated as priors for Monte-Carlo simulations. \\

Now we substantiate our work by presenting various figures in
form of plots. Fig.1a describes the Hubble parameter \d H(z) \d
versus redshift $z$ plot and error bar plot for the best fit
values of model parameters \g $H_{0}$, l and $\zeta$\g
using methods of least squire and minimum\g  $\chi^2$  \g
function value. Fig.1b  describes the Monte-Carlo Simulation-based
MCMC estimations related corner plots for model parameters  $H_{0}$, l and $\zeta$, whereas Fig.1c are step numbers plots for the model parameters showing that the simulations have been done properly.
%%%%%%%%%%%%%%%%%%%%%%%%%%%%%%%%%%%%%%%%%%%%%%%%%%%%%%%%%%
\begin{figure}[H]
	\centering
	
	a.\includegraphics[width=6cm,height=6cm,angle=0]{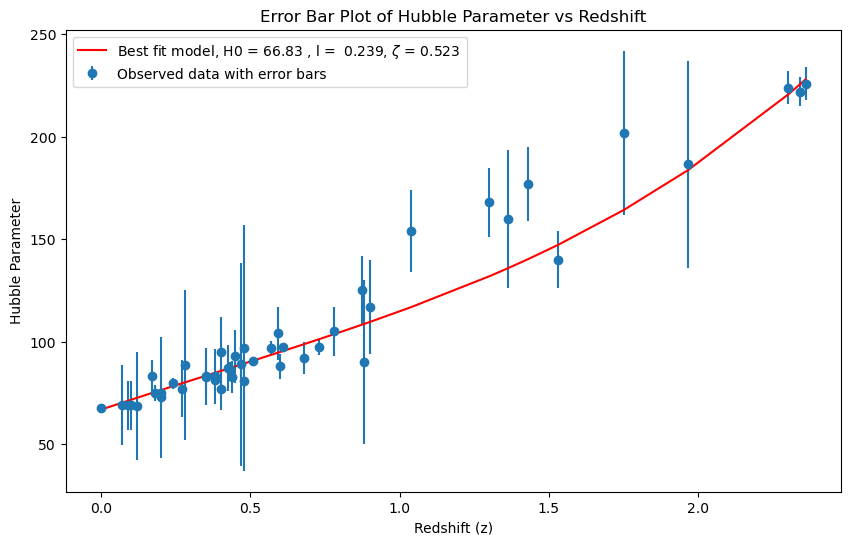}\\
    b.\includegraphics[width=6cm,height=6cm,angle=0]{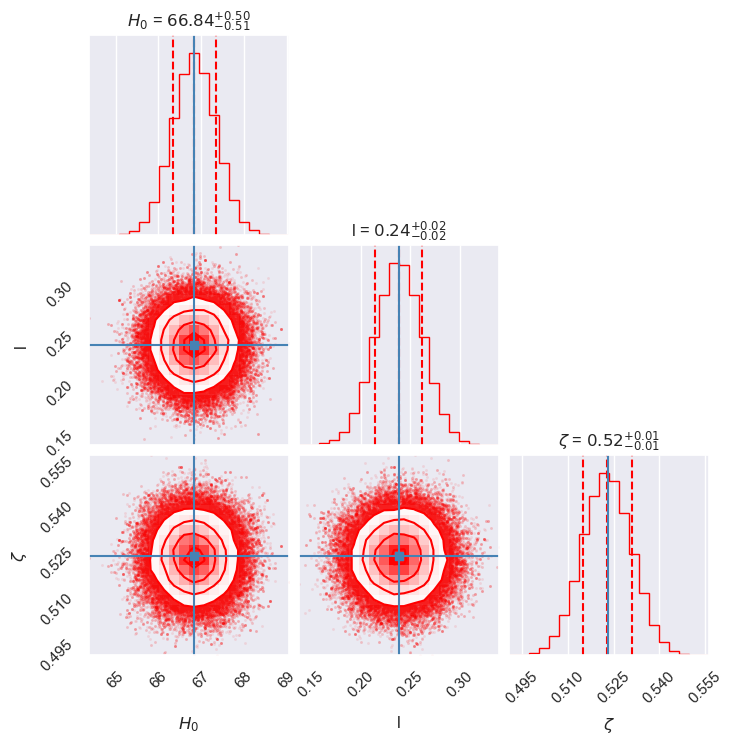}\\
    c.\includegraphics[width=6cm,height=6cm,angle=0]{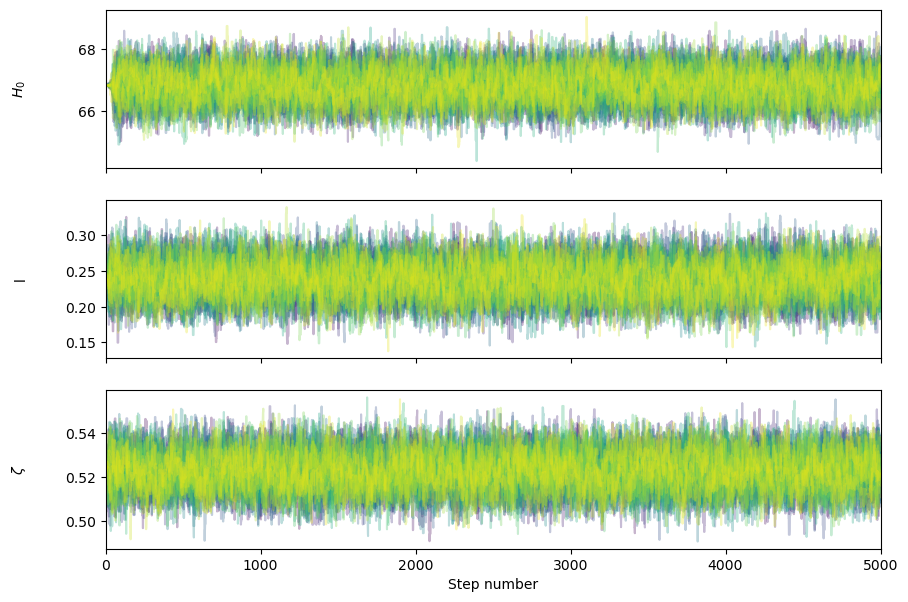}
	\caption{fig1a. Error bars plot for Hubble parameter. Parameters estimation based on minimum $\chi^{2}$.
	fig 1b and 1c are Corner and  Step numbers  plots for 46
	Hubble data set. Monte-Carlo Simulation-based MCMC estimations were carried out for Model parameters  $H_{0}$, l and $\zeta$. Estimated parameter values are described in the figures.}
\end{figure}\l{fig1}

%%%%%%%%%%%%%%%%%%%%%%%%%%%%%%%%%%%%%%%%%%%%%%%%%%%%%%%%%%%
\subsection{Estimations of Model parameters from Supernova SNIa Union 2.1 Compilation 680 data set.}\l{B}
The Supernova SNIa Union 2.1 Compilation data set\cite{refN42} is comprised
of 650 data sets of Distance modulus of SNIa supernovae for
various red shift in the range \d 0\leq z \leq 1.4\d associated
with observational errors. The theoretical formula for distance
modulus is given as:
\begin{equation}
	\mu_{th}(z;h,\boldsymbol{params}):= 5 \log\left[
	D_L(z;\boldsymbol{params})\right] +\mu_0(h) . \label{mu}
\end{equation}
Here $h$ is the Hubble constant in units of $100$ Km/s/
Mpc, $\boldsymbol{params}$ denotes the set of
cosmological parameters of interest other than $h$. In our model
params are \g $H_0$, l and $\zeta$.
\begin{equation}\no
	\mu_0(h):= 5\log\left(\frac{10^3 c/(\mbox{km/s})}{h}\right) =
	42.38-5\log h.
\end{equation}
and
\begin{equation}\no
	D_L(z;h,\boldsymbol{params})=(1+z)\int_0^z
	{\frac{dz'}{H(z';h,\boldsymbol{params})/H_0}}\;,
\end{equation}
is the luminosity distance. We use the Hubble function as per
Eq. (\r{11}). \\
Like in subsection (a), we use the union 2.1 data set to estimate
the model parameters \g  $H_{0}$, l and $\zeta$ \g in
multiple ways. First we fit distance modulus \g(\d m_{u}\d)\g
function given by Eq. (\r{mu}) to observational distance modulus
\g( $m_{u}$)\g data by method of least squire and estimates model
parameters. Thereafter assuming these estimated values as
initial guess, we use method of least $\chi^{2}$ to make more
refined estimations.
Finally, we carry Monte-Carlo simulations MCMC method to further
refine the estimations by assuming $\chi^{2}$ estimations as
initial guess. We recall that $\chi^{2}$ formula for distance
modulus \g(\d\mu\d)\g is as follows:
\be
\chi^{2}(H_{0}, l, \zeta ) =
\sum\limits_{i=1}^{580}\frac{(m_{bth}(z_{i},H_{0}, l, \zeta ) -
	m_{bob}(z_{i}))^{2}}{\sigma {(z_{i})}^{2}}.
\l{chimu}\ee	
Our estimations are shown in the following Table 2
\begin{table}[H] \l{table2}
	\bc
	{\begin{tabular}{c|c|c|c|}
			\hline\hline 
			Parameters & Least square & Least\g\d \chi^2\d\g & MCMC
			simulation\\
			& Estimations & Estimations & Estimations\\  
				\hline\\
		\d H_{0}\d & $69.826$ & $69.818\pm 0.23$ &
			$69.813 ^{+0.231} _ {-0.228}$\\
			\\
                \hline
			\\
			    l  & $ 0.698$ & $ 0.5498\pm 0.03$ & $0.550 \pm 0.030 $\\
			\\
			\hline
			\\
			\d\zeta\d & $0.802$ & $0.6956\pm0.01$ & $0.695 \pm 0.010 $\\
			\\
			
			\hline\hline
	\end{tabular}}
	\ec
	\caption{\small{ The best-fit values of the model parameters
			 $H_{0}$, l and $\zeta$ for the best fit distance modulus
			\g($\mu(z)$)\g curve.}}
\end{table}

Now we substantiate our work by presenting various figures in
the form of plots. Fig.2a describes the distance modulus \d \mu(z) \d
versus redshift $z$ Plot and Error bar Plot for the best fit
values of model parameters $H_{0}$, l and $\zeta$ \g
using methods of least squire and minimum\g  $\chi^2$  \g
function value. Fig.2b describes the Monte-Carlo Simulation-based
MCMC estimations related corner plots for Model parameters  $H_{0}$, l and $\zeta$.

\begin{figure}[H]
	\centering
	
	a.\includegraphics[width=6cm,height=6cm,angle=0]{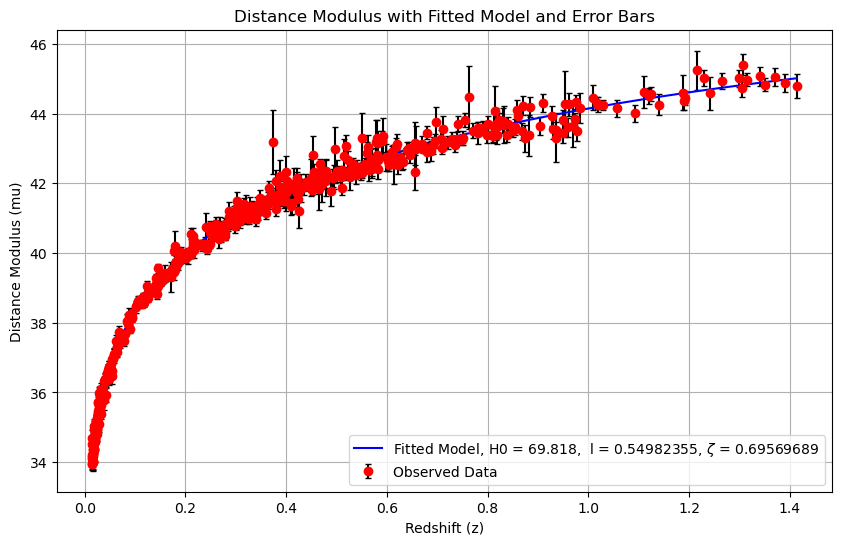}\\
	b.\includegraphics[width=6cm,height=6cm,angle=0]{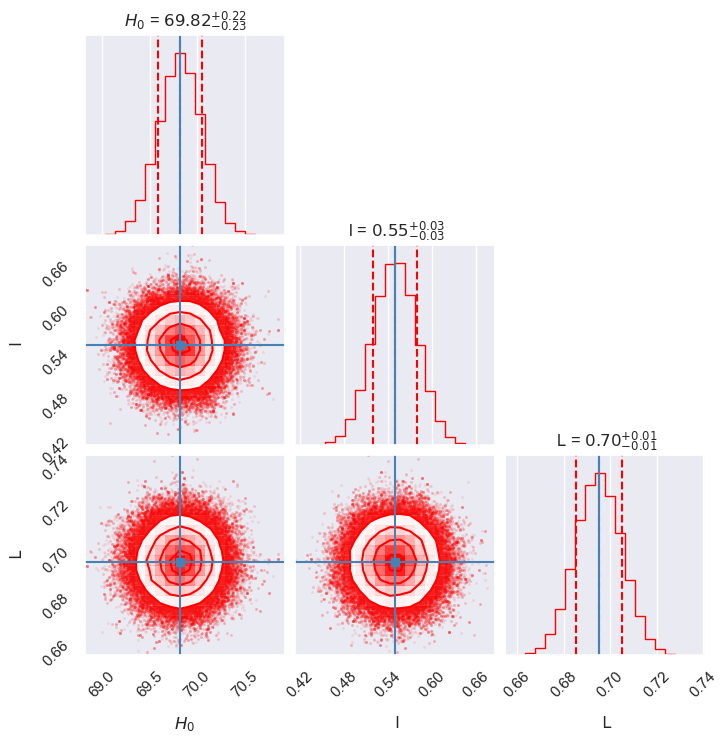}

	\caption{fig2a. Error bars plot for Distance modulus. Parameters estimation based on minimum $\chi^{2}$.
		fig2b  Monte-Carlo Simulation based MCMC estimations for Model parameters  $H_{0}$, l and $\zeta$. Corner plots for 518
		Distance modulus union 2.1 compilation data set. Estimated parameter values are described in the figures.}
\end{figure}\l{fig2}

%%%%%%%%%%%%%%%%%%%%%%%%%%%%%%%%%%%%%%%%%%%%%%%%%%%%%%%%%%%%%%%%
%%%%%%%%%%%%%%%%%%%%%%%%%%% Subsection C %%%%%%%%%%%%%%%%%%%%%%
\subsection{Estimations of Model parameters from Supernova SNIa
	1048 Pantheon data set.}\l{C}
The Supernova SNIa Pantheon data set\cite{refN43} is comprised of 1048 data
set of Apparent magnitude \g($m_b$)\g of SNIa supernovae for
various red shift in the range \g\d 0\leq z \leq 2.24\d \g,
associated with observational errors. The theoretical formula
for Apparent magnitude \g( $m_b$)\g is given as:

\begin{equation}\no
	m_b = M + \mu(z)
\end{equation}
where $ M $ is the absolute magnitude of the object and $ \mu $
is the distance modulus. We recall that Type Ia supernovae are
considered ``standard candles" because they have a relatively
uniform intrinsic brightness, which allows astronomers to use
them to measure distances accurately. The absolute magnitude of
a typical Type Ia supernova at its peak brightness is
approximately:
Absolute Magnitude: M = -19.09 in the B- band(blue light)
\begin{equation}
	m_{bth}(z;h,\boldsymbol{params}):= 5 \log\left[
	D_L(z;\boldsymbol{params})\right] + m_{b0}(h) .
	\label{mb}\end{equation}
Here $h$ is the Hubble constant in units of $100$ km/s/
Mpc, $\boldsymbol{params}$ denotes the set of
cosmological parameters of interest other than $h$, in our model
params are \g\d  l\g and \g\zeta \d.
\begin{equation}\no
	m_{b0}(h):= 5\log\left(\frac{10^3 c/(\mbox{km/s})}{h}\right)
	-19.09 = 23.29-5\log h.
\end{equation}
and
\begin{equation}\no
	D_{L}(z;h,\boldsymbol{params})=(1+z)\int_0^z
	{\frac{dz'}{H(z';h,\boldsymbol{params})/H_0}}\;,
\end{equation}
is the luminosity distance. We use the Hubble function as per
Eq. (\r{11}). \\
Like in subsections (a) and (b), we use the Pantheon data set to
estimate the model parameters \g $H_{0}$, l and $\zeta$ \g
in multiple ways. First we fit Apparent magnitude \g(\d m_b\d)\g
function given by Eq. (\r{mb}) to observational Apparent
magnitude \g(\d m_{b}\d)\g data by method of least squire and
estimates model parameters. Thereafter assuming these estimated
values as an initial guess, we use the method of least $\chi^{2}$ to
make more refined estimations.
Finally, we carry Monte-Carlo simulations MCMC method to further
refine the estimations by assuming $\chi^{2}$ estimations as
initial guess. We recall that $\chi^{2}$ formula for Apparent
magnitude \g(\d m_{b}\d)\g is as follows:
\be
\chi^{2}(H_0, l, \zeta) =
\sum\limits_{i=1}^{1048}\frac{(m_{b}th (z_{i},H_0, l, \zeta)
	- m_{b}ob (z_{i}))^{2}}{\sigma {(z_{i})}^{2}}.
\l{chimb}\ee	
Our estimations are shown in the following Table 3
%%%%%%%%%%%%%%%%%%%%%%%%%%%%%%%% Table IV %%%%%%%%%%%%%%%%%%%%%%%
\begin{table}[H]\l{table3}
	\centering  
	{\begin{tabular}{c|c|c|c|} 
			\hline\hline 
			Parameters & Least square & Least\g\d \chi^2\d\g & MCMC
			simulation\\
			& Estimations & Estimations & Estimations\\ 
			\hline\\ 
			 \d H_{0}\d & $79.143$ & $79.148\pm 0.21$ &
			$79.149 ^{+0.208} _ {-0.211}$\\
			\\
                \hline
			\\
			    l  & $0.675$ & $0.669\pm 0.03$ & $0.670 \pm 0.030$\\
			\\
			\hline
			\\
			\d\zeta\d & $ 0.763$ & $ 0.749\pm0.01$ & $0.748 \pm0.01$\\
			\\
		    \hline\hline
	\end{tabular}}
	\caption{\small{ The best-fit values of the model parameters
			 $H_{0}$, l and $\zeta$ for the best fit Apparent magnitude
			\g(\d m_{b}(z)\d)\g curve.}}
\end{table}
%%%%%%%%%%%%%%%%%%%%%%%%%%%%%%%%%%%%%%%%%%%%%%%%%%%%%%%%%%%%%%%%%
Now we substantiate our work by presenting various figures in
form of plots. Fig.3a describes the distance modulus \d \mu(z) \d
\g versus redshift $z$\g Plot and Error bar Plot for the best fit
values of model parameters \g$H_{0}$, l and $\zeta$ \g
using methods of least squire and minimum\g \d \chi^2 \d \g
function value. Fig.3b describes the Monte-Carlo Simulation-based
MCMC estimations for Model parameters $H_{0}$, l and $\zeta$. 
%%%%%%%%%%%%%%%%%%%%%%%%%%%%%%%%%%%%% Figures 7(a) and 7(b) %%%%%%%%%%%%%
\begin{figure}[H]
	\centering
	
	a.\includegraphics[width=6cm,height=6cm,angle=0]{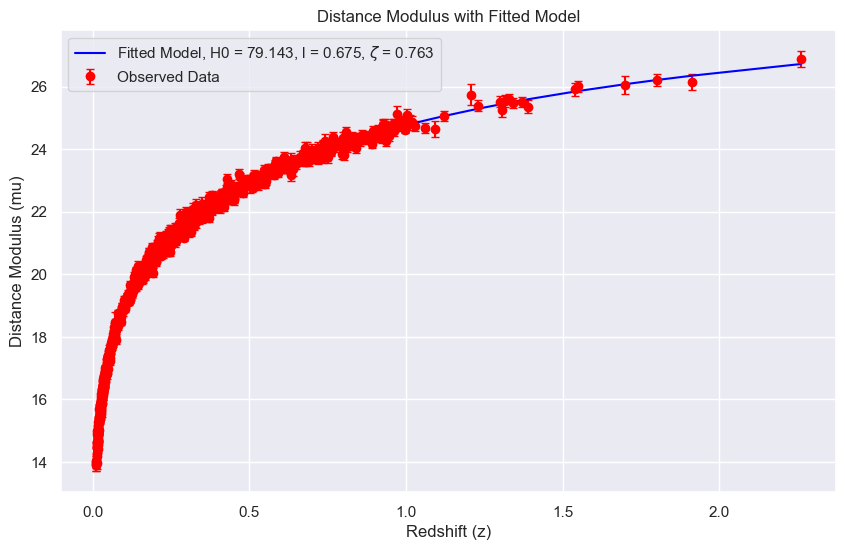}\\
	b.\includegraphics[width=6cm,height=6cm,angle=0]{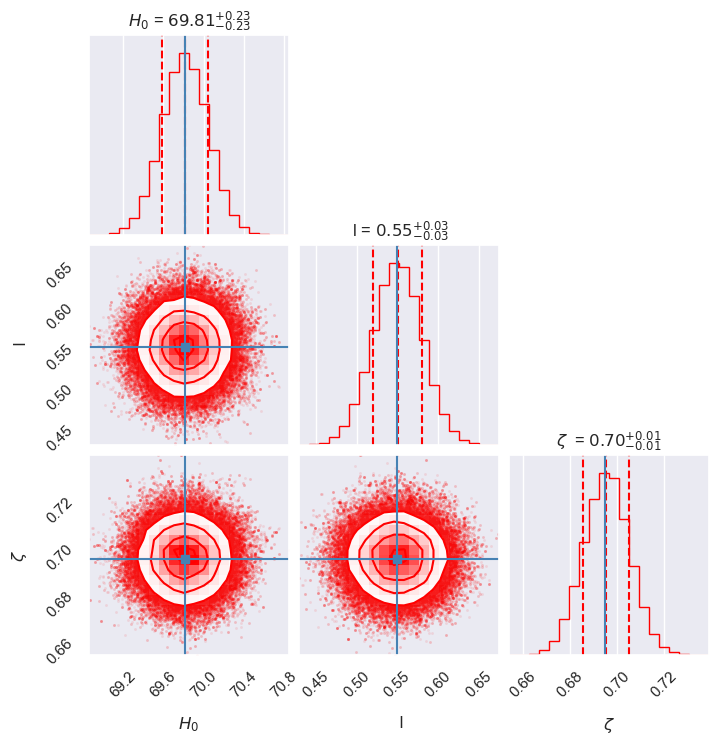}
	
	\caption{fig3a. Error bars plot for Apparent magnitude. Parameters estimation based 
        on minimum $\chi^{2}$.
		fig3b  Monte-Carlo Simulation based MCMC estimations for Model parameters  $H_{0}$, l and $\zeta$. Corner plots for 1048 Apparent magnitude Pantheon data set. Estimated parameter values are described in the figures.}
\end{figure}\l{fig3}

%%%%%%%%%%%%%%%%%%%%%%%%%%%%%%%%%%%%%%%%%%%%%%%%%%%%%%%%%%%%%%%%%%%
%%%%%%%%%%%%%%%%%%%%%%%%%%%%%%%%% Subsection D %%%%%%%%%%%%%%%%%%%%%
\subsection{Baryon Acoustic Oscillations Analysis:}\l{D} BAO
refers to the regular, periodic fluctuations in the density of
visible baryonic matter (normal matter) of the universe, caused
by acoustic waves in the early universe. BAO provides a
"standard ruler" for the length scale in cosmology. The scale of BAO
is determined by the sound horizon at the time of recombination.
By measuring the scale of BAO in the distribution of galaxies or
the Lyman-alpha forest of quasars, astronomers can determine
distances across different epochs of the universe. BAO
measurements help to trace the expansion history of the
universe, complementing the information obtained from SNIa and
CMB.
%%%%%%%%%%%%%%%%%%%%%%%%%%%%% Subsection 1 %%%%%%%%%%%%%%%%%%%%%%%%%%%%%%%%
\subsubsection{Key Parameters in BAO Analysis}
\textbf{Sound Horizon ( $ r_s(z) $):} The distance that sound
waves could travel in the early universe before recombination.
This serves as the standard ruler for BAO measurements.It is
defined as $$r_s(a) = \int_0^a \frac{c_s da}{ a^2 H(a) }. $$
It's typically about 150 megaparsecs (Mpc).\\

\textbf{Angular Diameter Distance ( $ d_A(z) $):} The distance
derived from the angular size of BAO features at a given
redshift. It is defined as
$$  d_A(z) = c \int_0^{z} \frac{dz'}{H(z')} $$.
\textbf{Volume-Averaged Distance ($ D_v(z) $):} A combination of
the angular diameter distance and the Hubble parameter, given
by: $$D_v(z) = \bigg(\frac{z {d_A}^2 (z)}{ H(z)}
\bigg)^{\frac{1}{3}} $$
\textbf{Ratio of Distances:} The distance redshift ratio is
often used in BAO analysis to compare with theoretical
predictions. It is is given by:
\begin{equation}
	d_z(z)= \frac{r_s(z*)}{D_v(z)}
\end{equation}
where $ r_s(z*) $ denotes the co-moving sound horizon at the
time when photons decouple and $ z* $ is the photons decoupling
redshift. We consider $ z* = 1090 $ for the analysis.
In this work, we consider a sample of BAO distance measurements
from different surveys such as SDSS(R)\cite{refN44}, the
6d F Galaxy survey \cite{refN45}, BOSS CMASS \cite{refN46},
and three parallel measurements from the Wiggle Z survey
\cite{refN47,refN48,refN49,refN50}.\\
We consider the following observational data set \cite{refN51} for our
analysis which are described as follows:\\
\d z_{BAO} \d = ([0.106, 0.2, 0.35, 0.44, 0.6, 0.73])\\
\d d_z(z_{BAO})\d  = ([30.95, 17.55, 10.11, 8.44, 6.69, 5.45])\\	
\d\sigma\d  = ([1.46, 0.60, 0.37, 0.67, 0.33, 0.31])\\
The $ \chi^{2}_{BAO} $ corresponding to BAO measurements is
given by \cite{refN51}
\begin{equation}\l{chibao}
	\chi_{BAO}^2(\a,\b,\c,H_0) = X^{T} C^{-1} X,
\end{equation}
where,
\begin{equation}\l{14}
	X = 
	\begin{bmatrix}
		$$ \frac{d_A(z*)}{D_v(0.106)}-30.84$$ \\
		$$ \frac{d_A(z*)}{D_v(0.35)}- 10.33 $$\\
		$$ \frac{d_A(z*)}{D_v(0.57)}- 6.72 $$\\
		$$ \frac{d_A(z*)}{D_v(0.44)}- 8.41 $$\\
		$$ \frac{d_A(z*)}{D_v(0.6)}- 6.66 $$\\
		$$ \frac{d_A(z*)}{D_v(0.73)}- 5.43 $$
	\end{bmatrix}
\end{equation}

and $ C^{-1} $ is the inverse of the covariance matrix given by

\begin{equation}\l{15}
C^{-1} = 
\begin{bmatrix}
 0.52552 & -0.03548 & -0.07733 & -0.00167 & -0.00532 & -0.0059 \\
 -0.03548 & 24.9707 & -1.25461 & -0.02704 & -0.08633 & -0.09579 \\
 -0.07733 & -1.25461 & 82.9295 & -0.05895 & -0.18819 & -0.20881 \\
 -0.00167 & -0.02704 & -0.05895 & 2.9115 & -2.98873 & 1.43206 \\
 -0.00532 & -0.08633 & -0.18819 & -2.98873 & 15.9683 & -7.70636 \\
 -0.0059 & -0.09579 & -0.20881 & 1.43206 & -7.70636 & 15.2814 \\
\end{bmatrix}
\end{equation}
We explain that the ratio \g\d d_z(z)\d\g is in fact function of
model parameters \g $H_{0}$, l and $\zeta$\g. The same is
true for \g $\chi^2$\g function also. So we use it to estimate
our model parameters as we did in previous sections. We present our results in the form of Table 4 and figures.
%%%%%%%%%%%%%%%%%%%%%%%%%%%%%%%%%%%%%% Table IV %%%%%%%%%%%%%%%%%%%%%%%%%%
\begin{table}[H] 
	\centering
	{\begin{tabular}{c|c|c|c|}
			
			\hline\hline 
			Parameters & Least square & Least\g\d \chi^2\d\g & MCMC
			simulation\\
			& Estimations & Estimations & Estimations\\\\
			\hline\\
			\d H_{0}\d & $69.587$ & $71.48\pm 3.93$ &
			$71.363^ {+3.907} _{-3.897}$\\
			\\
                \hline
			\\
			    l  & $0.0459$ & $0.033\pm 0.02$ & $0.033 \pm 0.019$\\
			\\
			\hline
			\\
			\d\zeta\d & $0.974$ & $0.964\pm0.02$ & $0.964\pm 0.018$\\
			\\
			
			\hline\hline
	\end{tabular}}
	\caption{\small{ The best-fit values of the model parameters
			 $H_{0}$, l and $\zeta$ for the best fit distance ratio
			\g\d d_{z}\d\g curve. }}
\end{table}\l{table4}
%%%%%%%%%%%%%%%%%%%%%%%%%%%%%%%%%%%%%%%%%%%%%%%%%%%%%%%%%%%%%%%%%%%%
%%%%%%%%%%%%%%%%%%%%%%%%%%%% Figure 4(a) and 4(b) %%%%%%%%%%%%%%%
\begin{figure}[H]
	\centering
	
		a.\includegraphics[width=6cm,height=6cm,angle=0]{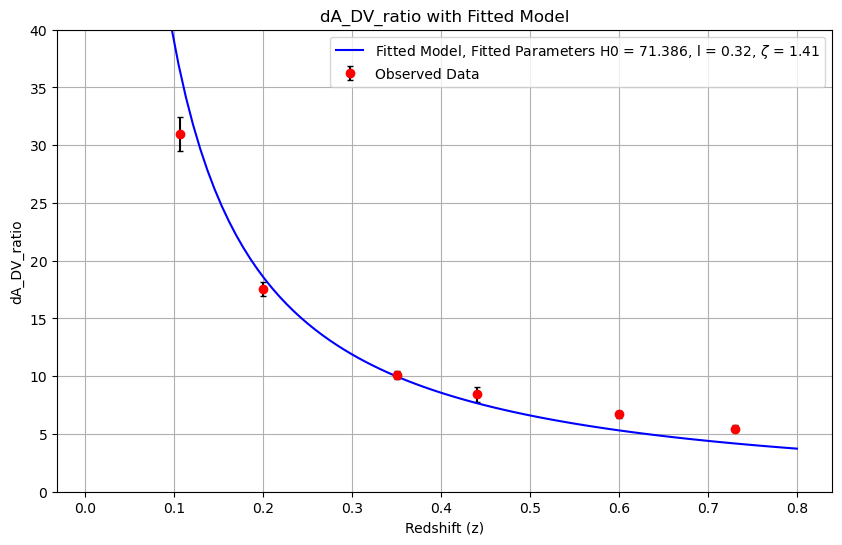}
		b.\includegraphics[width=6cm,height=6cm,angle=0]{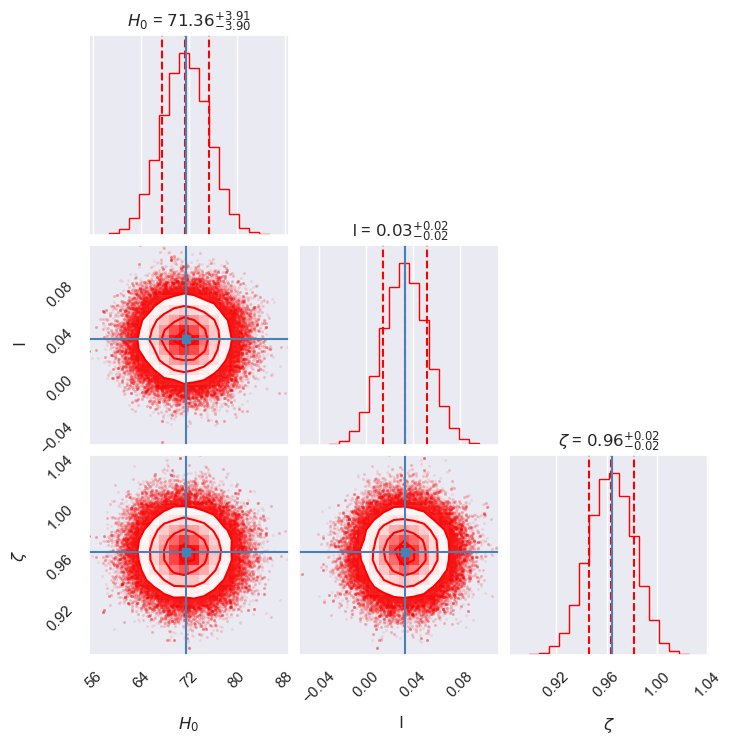}
	\caption{fig4a Error bars plot  for the distance ratio \g\d d_{z}\d\g versus red shift. Parameters estimation based on minimum $\chi^{2}$.
		fig4b  Monte-Carlo Simulation based MCMC estimations for Model parameters  $H_{0}$, l and $\zeta$. Corner plots for 6
	 distance ratio \g\d d_{z}\d\g data set with standard errors. Estimated parameter values are described in the figures.}
	
\end{figure}\l{fig4}

%%%%%%%%%%%%%%%%%%%%%%%%%%%%%%%%%%%%% Subsection E %%%%%%%%%%%%%%%%%%%%%%
\subsection{Estimations of Model parameters from the combined
	data sets out of OHD, BAO, Pantheon, and Union2.1 Compilation
	:}\l{E}
In this section, we continue our estimations of model parameters
 $H_{0}$, l and $\zeta$ with the help of combined data sets
formed out of OHD, BAO, Pantheon, and Union2.1 Compilation. For
this we construct a combined $\chi^{2}$ function by adding
individual $\chi^{2}$ function. For example, suppose we want to
combine OHD and BAO data sets then our combined $\chi^{2}$ function will be as follows:
\be
\chi^2_{OHD + BAO}(H_0, l, \zeta) = \chi^2_{OHD}(H_0, l, \zeta )+ \chi^2_{PAN}(H_0, l, \zeta)+ \chi^2_{BAO}(H_0, l, \zeta )
\ee\l{chicomb}
By minimizing the above function by giving proper initial values
and ranges of model parameters, we estimate model parameters for
combined data sets(CDS). We present the following Table 5
which displays estimated results for various CDS.

%%%%%%%%%%%%%%%%%%%%%%%%%%%%%%%%%%%%%%%% Table VI %%%%%%%%%%%%%%%%%%%%%%%%%%%%
\begin{table}[H]
	\begin{center}
		\begin{tabular}{l c c c c c} 
			\hline 
			\hline 
			
			\\ 
			Parameters &~~OHD+BAO &   ~OHD+Pan+BAO+Union
			\\
			\hline
			\\  
			
			\d H_{0}\d & $66.912 ^ {+0.497}_{-0.501}$ &
			$74.216^{+0.150}_{-0.148}$\\
			\\
                \hline
			\\
			    l   & $0.112 \pm 0.015$ & \d0.276 \pm 0.012\d   \\
			\\
			\hline
			\\
			\d\zeta\d  & $0.596 \pm 0.007$  &\d 0.661 \pm 0.005 \d  \\
			\\
			\\
			\hline\hline   
		\end{tabular}  
		\caption{Estimated values of Model Parameters  $H_{0}$, l and $\zeta$}
	\end{center}
\end{table}\l{table5}
%%%%%%%%%%%%%%%%%%%%%%%%%%%%%%%%%%%%%%%%%%%%%%%%%%%%%%%%%%%%%%%%%%%%%%%
Note that the above-estimated values are based on carrying
Morkov chain Monte-Carlo simulations which further refine the
estimations by minimum $\chi^{2}$.
Now, we present the following Corner plots and Step number plots
to show the results of MCMC simulations. The details have been
given in the caption of each plot.

%%%%%%%%%%%%%%%%%%%%%%%%%%%%%%% Figure 13 %%%%%%%%%%%%%%%%%%%%%%%%%%%%%%%

\begin{figure}[H]
	\centering
		\includegraphics[width=4cm,height=6cm,angle=0]{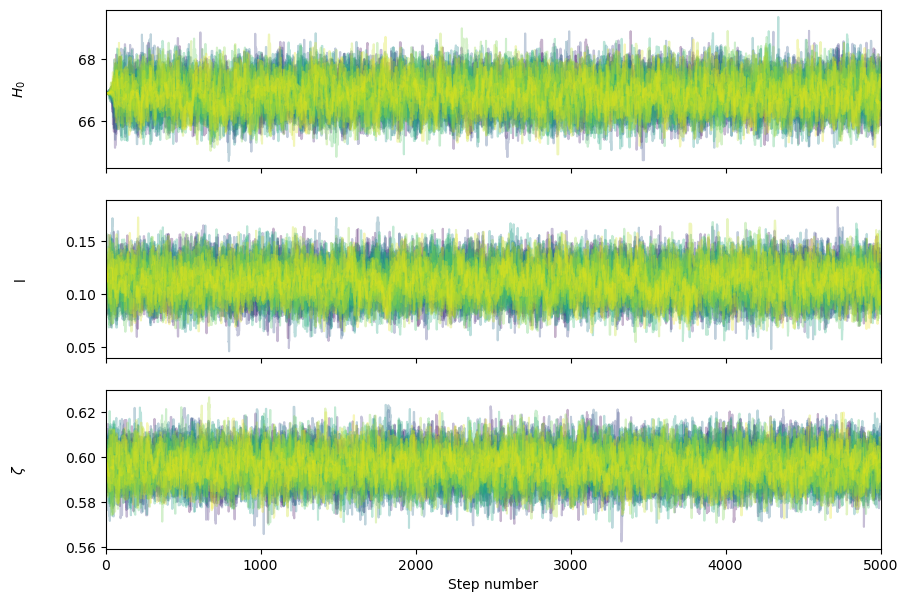}
		\includegraphics[width=4cm,height=6cm,angle=0]{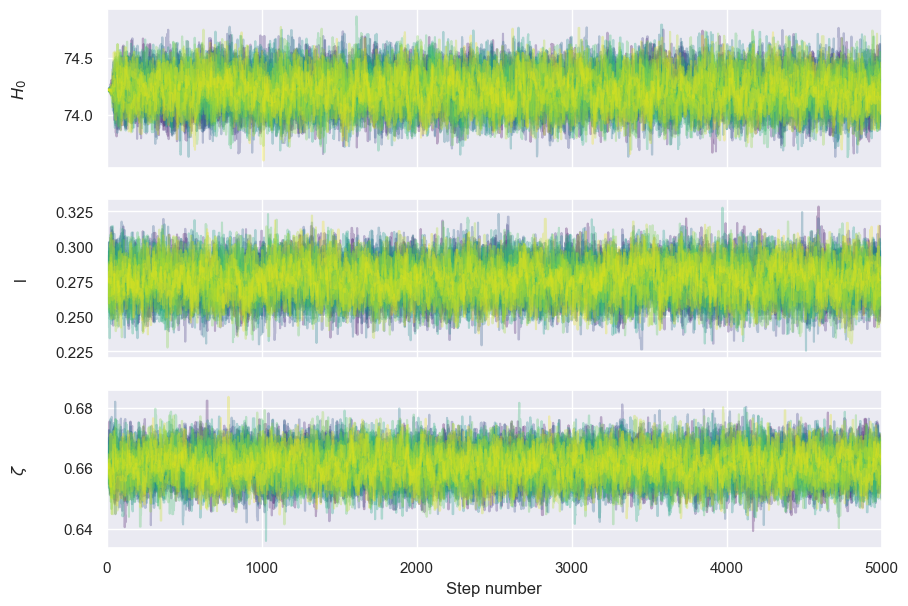}
		\includegraphics[width=4cm,height=6cm,angle=0]{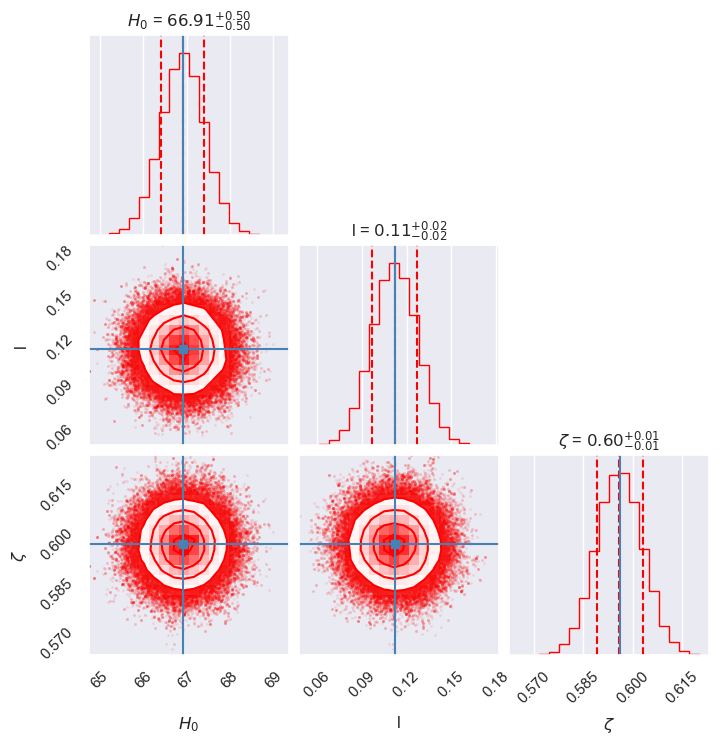}
		\includegraphics[width=4cm,height=6cm,angle=0]{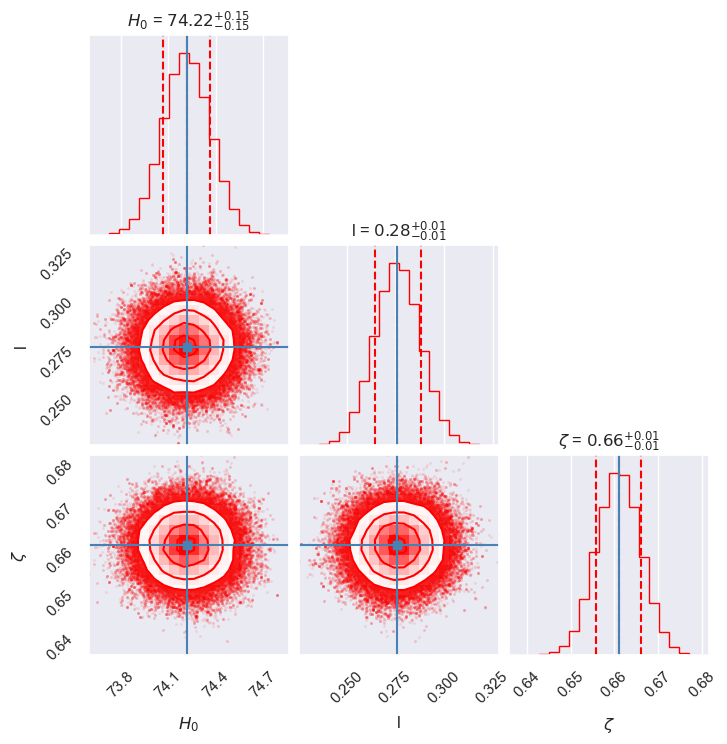} 
	
	\caption{\small{Corner plots for Combined Hubble plus BAO plus
			Union 2.1 and combined Hubble plus Union2.1  plus Bao plus Pantheon data sets, based on Markov Chain Monte-Carlo(MCMC) Simulation
			estimations of Model parameters $H_{0}$, l and $\zeta$ .}
		The first two figures are Step numbers versus Model parameters
		 plots for combined data sets
		described earlier. }
\end{figure}\l{fig5}
%%%%%%%%%%%%%%%%%%%%%%%%%%%%
The current estimates of the Hubble constant ($H_{0}$), which describes the rate of the universe's expansion, generally fall within two primary ranges depending on the measurement technique. The Hubble Key Project team in 2001 found that $H_0 = 72 \pm 3$ (statistical) $\pm 7$ (systematic) was the result of a Cepheid-based calibration \cite{refN50a}. After two more decades of work with HST, Spitzer, and numerous other ground-based telescopes, measurements of $H_{0}$ improved, with estimated accuracies today falling between 2 and 5\% \cite{refN50b}.
 While observations utilizing the tip of the red giant branch (TRGB) \cite{refN50c,refN50d} produce slightly lower values, closer to 70 Km/s/Mpc, the Cepheid calibration of $H_{0}$ \cite{refN50e,refN50f} continues to yield values of $H_{0}=73.04 \pm 1.04$ Km/s/Mpc. The precision of recent estimates of $H_{0}$ from CMB observations is very enormous; the Planck spacecraft \cite{refN50g} produced $H_{0} = 67.4 \pm 0.54$ Km/s/Mpc (better than 1\%). Megamaser Cosmology Project \cite{refN50h} finds $H_{0} = 73.9 \pm 3.0$ Km/s/Mpc, and H0LiCOW \cite{refN50i} finds $H_{0} = 73.3^{+1.7}_{-1.8}$Km/s/Mpc. The gap between these ranges, known as the ``Hubble tension," suggests potential new physics or systematic errors in one or both measurement approaches. Although local measurements occasionally hint at higher values, exceeding 79 Km/s/Mpc is considered unlikely under the current framework. Current efforts focus on refining methods, reducing uncertainties, and exploring alternative models that might reconcile the tension \cite{refN50h,refN50i}. Thus, while the idea of $H_{0}$ exceeding 79 Km/s/Mpc is not entirely excluded, it would imply significant adjustments to our cosmological understanding.
 
%%%%%%%%%%%%%%%%%%%%%%%%%%%%%%%%%%%
%%%%%%%%%%%%%%%%%%%%%%%%%%%%%%%%%%%% Section 4 %%%%%%%%%%%%%%%%%%%%%%%%%%%%%%%%%%%%%%%%%%%

	\section{Conversion of Redshift into Time and Age of the
		Universe:}\l{section4}
	We may get time in a billion years with the help of redshift from
	the following relation :
	\be\no
	\frac{dz}{dt} = -(1+z)H(z)
	\ee
	This equation is integrated to yield the following relationship
	between elapsed time\g$(t_0- t_z$)\g from present and redshift
	$z$:
	
	\begin{equation}\label{16}
t_{0}-t=\int_0^z \frac{3 \sqrt{\zeta +1}}{(z+1)^{5/2} (z+1)^{-\frac{3 \zeta }{2}} \sqrt{l^2 \left((z+1)^{3 \zeta +3}-1\right)+9 (\zeta +1)}} \, dz.
\end{equation}
	
	where $t_0$ is present time at $z=0$ and $t_z$ is the time at
	redshift $z$, so that $(t_0 -t_z$) will be the elapsed time
	from the present at the redshift $z$.
	We note that unit of Hubble parameter $H(z)$ \small{ Km/sec/Mps}
	which is a unit of reciprocal of time. We use the following
	conversion formula to express unit $\frac{Mps}{km/sec}$ of the reciprocal of Hubble parameter H(z)  in time i.e. billion yrs.
	\be\no
	\frac{Mps}{Km/sec} = \small{976.32}\g \small{billion ~ yrs}.
	\ee
	
The universe's present age is represented as $ t_{0}. $ It is written as
\begin{equation}\label{17}
t_{0}=\lim_{x \to \infty}\int_0^x \frac{3 \sqrt{\zeta +1}}{(z+1)^{5/2} (z+1)^{-\frac{3 \zeta }{2}} \sqrt{l^2 \left((z+1)^{3 \zeta +3}-1\right)+9 (\zeta +1)}} \, dz
\end{equation}
Integrating Eq. (\ref{17}), we get
\ba
H_{0}  t_{0} &=& 0.940404\g for \g H_{0}=  74.216\g l = 0.276,\g and \g \zeta = 0.661.\l{18}\\
              &=& 1.05451\g for \g  H_{0}= 66.912\g l = 0.112,\g and \g \zeta = 0.596\l{19}.
\ea

Therefore, the present age of the universe for the derived model is estimated as \ba
  t_{0} &=& 12.3738\g  Gyrs  \g for \g H_{0}=  74.216\g l = 0.276,\g and \g \zeta = 0.661.\l{20}\\
              &=& 15.3864\g  Gyrs  \g for \g  H_{0}= 66.912\g l = 0.112,\g and \g \zeta = 0.596\l{21}.
\ea

Figure 6 shows the variation of $ H_{0} (t_{0}-t) $ over redshift $ z $ at different parameters based on our combined data sets.
According to WMAP data, the empirical value of the universe's current age is $ t_{0}=13.73^{+.13}_{-.17} $ $ Gyrs $.

%%%%%%%%%%%%%%%%%%%%%%%%%%%%%%%%%%%%%%%%%% Figure 4 %%%%%%%%%%%%%%%%%%%%%%%%%%%%%%%%%
\begin{figure}[H]
\centering
	\includegraphics[scale=0.9]{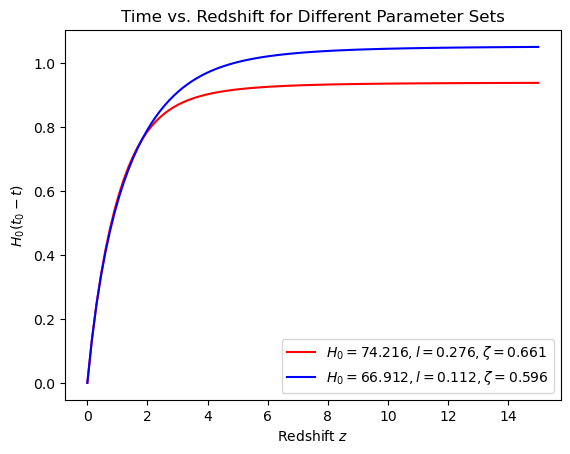}
\caption{Plot of $ H_{0} (t_{0}-t) $ versus $ z $. }	
\end{figure}\l{fig6}
%%%%%%%%%%%%%%%%%%%%%%%%%%%%%%%%%%%%%%%%%%%%%%%%%%%%%%%%%%%%%%%%%%%%%%%%%%%%%%%%%%%%%%%%
\section{Density of the universe:}
The density of the universe is obtained from Eqs.(\r{10b}) and (\r{11}) as:

\begin{equation}
  \rho = \rho_c \(\frac{ (z+1)^{3} (z+1)^{-3 \zeta } ({l}^2 \left((z+1)^{3 \zeta +3}-1\right)+9 (\zeta +1))}{9 (\zeta +1)} -\frac{ l^{2}}{9(1+z)^{6}}\) \l{22} 
\end{equation}
where $\rho_c= \frac{3 c^2 H_0^2}{8\pi G}$ is the critical density of
the universe.\\
 We present the following Figure to show the variation of $ \rho/\rho_c $ over red shift `z'. 
 \begin{figure}[H]
\centering
	\includegraphics[scale=0.9]{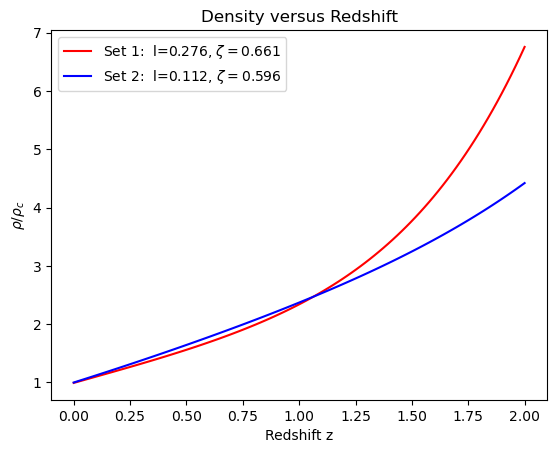}
\caption{Plot of $ \rho/\rho_c $ versus $ z $. }
\end{figure}\l{fig7} 
        The plot is an increasing one, This shows that in the
	past the density of the universe was more and it is decreasing
	over time. We note that the red shift is proportional to the time.
       We note that $ \frac{3c^2}{8 \pi G} =0.000189*10^{-29} g/cm^3 .$ The density at present time(z=0) is obtained as
       \ba
       \rho_0  &=& 4.06629*10^{-31} gm/cm^3\g for \g H_{0}=  74.216\g, l = 0.276,\g and \g \zeta = 0.661.\l{23}\\
              &=& 8.45009*10^{-30} gm/cm^3\g for \g  H_{0}= 66.912\g, l = 0.112,\g and \g \zeta = 0.596\l{24}.
        \ea      
  The current density of the universe is a critical parameter in cosmology, encapsulated by the concept of the critical density, which is about $9.2\times 10^{-30}g/cm^{3}$. This value corresponds to the density required to create a flat geometry for the universe under the Lambda Cold Dark Matter ($\Lambda$CDM) model. \\
  The universe's density can be broken into components:
  (i) Dark Energy: Constitutes roughly 68-70\% of the total density, driving the universe's accelerated expansion.
  (ii) Dark Matter: Accounts for about 25-27\%, comprising non-luminous and non-baryonic matter.
  (iii) Baryonic Matter (ordinary matter): Makes up approximately 4-5\%, including stars, planets, and interstellar gas. 
  These proportions are supported by recent research from the Dark Energy Survey (DES) and the Dark Energy Spectroscopic Instrument (DESI), but early-universe data (such as cosmic microwave background measurements) and late-universe probes (such as supernovae and galaxy distributions) appear to differ slightly in terms of parameters. 
  For example, DESI has offered high-precision data that support the fundamental ($\Lambda$CDM) model while suggesting that dark energy dynamics and matter distribution may be improved throughout cosmic time.

%%%%%%%%%%%%%%%%%%%%%%%%%%%%%%%%%%%% Section 5 %%%%%%%%%%%%%%%%%%%%%%%%%%%%%%%%%%%%%
\section{Deceleration Parameter}
Expression for the deceleration parameter $q(z)$ as a function of 'z' is obtained by using Eqs.(\ref{11}) and (\ref{12}) as follows:
\be\l{25}
q(z)=\frac{3 \left(\text{l}^2 \left( +2 (z+1)^{3\zeta +3}-1\right)-9 \zeta^2+9\right)}{2 \left(\text{l}^2 \left((z+1)^{3\zeta +3}-1\right)+9 (\zeta +1)\right)}-1
\ee
%%%%%%%%%%%%%%%%%%%%%%%%%%%%%%%%%%%%%%%%%%%%%%%%%%%%%
%\begin{figure}[H]
%\centering
%	\includegraphics[scale=0.9]{dec para.png}
%\caption{Plot of $ q $ versus $ z $. }	
%\end{figure}\l{fig8}
%%%%%%%%%%%%%%%%%%%%%%%%%%%%%%%%%%%%%%%%%%%%%%%%
%%%%%%%%%%%%%%%%%%%%%%%%%%%%%%%%%%%%%%%%%%%%%%%%%%%%%
\begin{figure}[H]
\centering
	\includegraphics[scale=0.9]{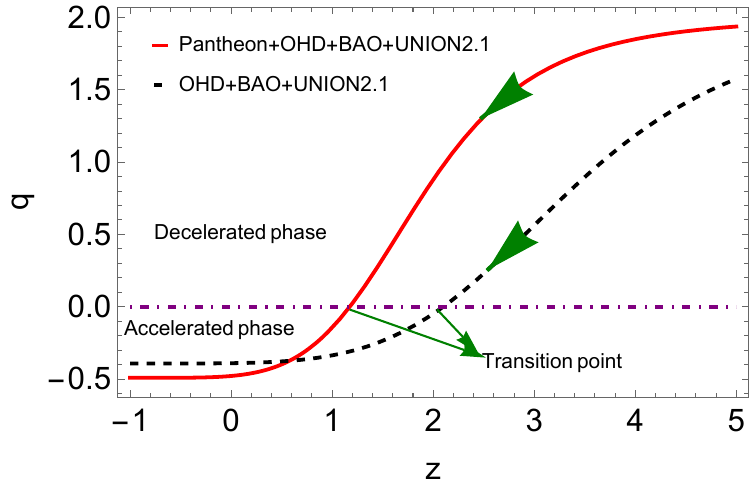}
\caption{Plot of $ q $ versus $ z $. }	
\end{figure}\l{fig9}
%%%%%%%%%%%%%%%%%%%%%%%%%%%%%%%%%%%%%%%%%%%%%%%%%%%%
Figures 8 describes the dynamics of deceleration parameter $q$ for redshift $z$
corresponding to the two sets of parameters. One with \d\g H_{0}=  74.216,\g l = 0.276,\g  and \g \zeta = 0.661,\d  and the other with \d H_{0}= 66.912,\g l = 0.112,\g  and \g \zeta = 0.596 \d. We recall that these parameter values were obtained based on Combined Hubble plus BAO plus Union 2.1 and combined Hubble plus Union 2.1 plus BAO plus Pantheon data sets. We obtain the present values of the deceleration parameter $q_{0}$ as
\ba
q_0 &=& -0.478804\g for \g H_{0}=  74.216\g, l = 0.276,\g and \g \zeta = 0.661.\l{26}\\
              &=& -0.391909\g for \g  H_{0}= 66.912\g, l = 0.112,\g and \g \zeta = 0.596\l{27}.
\ea
The above obtained results are in good agreement with those reported in \cite{refN50j,refN50k,refN50l}.

We also obtain the transition redshifts $z_t $ for the deceleration parameter when the universe enters the accelerating phase after passing through the decelerating one. It is described as:
\ba
 z_t &=& 1.1745\g for \g H_{0}=  74.216\g l = 0.276,\g and \g \zeta = 0.661.\l{28}\\
              &=&  2.1004\g for \g  H_{0}= 66.912\g l = 0.112,\g and \g \zeta = 0.596\l{29}.
\ea
The aforementioned findings are consistent with those found in \cite{refN50j,refN50k,refN50l,refN50m,refN50n}.

%%%%%%%%%%%%%%%%%%%%%%%%%%%%%%%%%%%% Section 6 %%%%%%%%%%%%%%%%%%%%%%%%%
\section{Statefinders}

In this section, we have discussed the statefinder diagnostic. Two well-known geometrical factors that describe the expansion history of the universe are the Hubble parameter $H$, which represents the expansion rate of the universe, and the deceleration parameter $q$, which indicates the rate of acceleration/deceleration of the expanding cosmos.  Only the scale factor $a$ affects these parameters. But with the proliferation of cosmological models and the notable improvement in cosmology observational data accuracy, these two parameters are no longer sensitive enough to distinguish between theories.
In order to differentiate between a growing variety of cosmological models that contain dark energy, the statefinder diagnostic has been developed. The statefinder diagnostic is probably a useful tool for differentiating cosmological models since they all have distinct evolutionary trajectories in the $(r - ~s)$ plane. The striking characteristic is that the $\Lambda CDM$ model, shown in figure 10a, corresponds to `$(r-~~ s) = (1,0)$'.
 \\
Numerous dark energy models, such as the holographic dark energy models, the phantom, the quintessence, the Chaplygin gas, and the interacting DE models, have been examined in previous researches \cite{ref102N52,ref104N54,ref105N55,ref106N56}. In the $(r - ~s)$ plane, one can determine how a particular cosmological model differs from a $\Lambda$CDM model.

\begin{equation}\label{29}
r= \frac{\ddot{H}}{H^{3}}+3\frac{\dot{H}}{H^2}+1.
\end{equation}
\begin{equation}\label{30}
s= \frac{r-1}{3 (q-\frac{1}{2})}.
\end{equation}
It is important to note that, according to \cite{ref102N52,ref107N57,ref108N58}, different combinations of $r$ and $s$ indicate different DE models.
Figure 9(a) shows that our model matches the $\Lambda$CDM $(r = 1, s = 0)$ point and is located in the quintessence area $(r < 1, s > 0)$ and the Chaplygin gas region $(r > 1, s < 0)$. In addition, the evolutionary trajectories in the $(r-q)$ plane are displayed in figure 9(b). The SS model (desitter expansion) is depicted in the image at the fixed point $(q =-1, r=1)$. According to the study's findings, the $q-r$ plane's evolutionary paths start beyond the SCDM ( $r = 1$, $q = 0.5$), or the matter-dominated Universe of the past, and converge below the $\Lambda$CDM fixed point ($q = -0.5$, $r = 1$) in the universe's late time evaluation. The recent phase shift of the universe is also justified by the fact that $q$ changes sign from +ve to -ve.

%%%%%%%%%%%%%%%%%%%%%% Fig 6a & Fig 6b %%%%%%%%%%%%%%%%%%%%%%%%%%%%%%%%%%%%%%%%%%%
\begin{figure}[H]
\centering
	(a)	\includegraphics[scale=0.80]{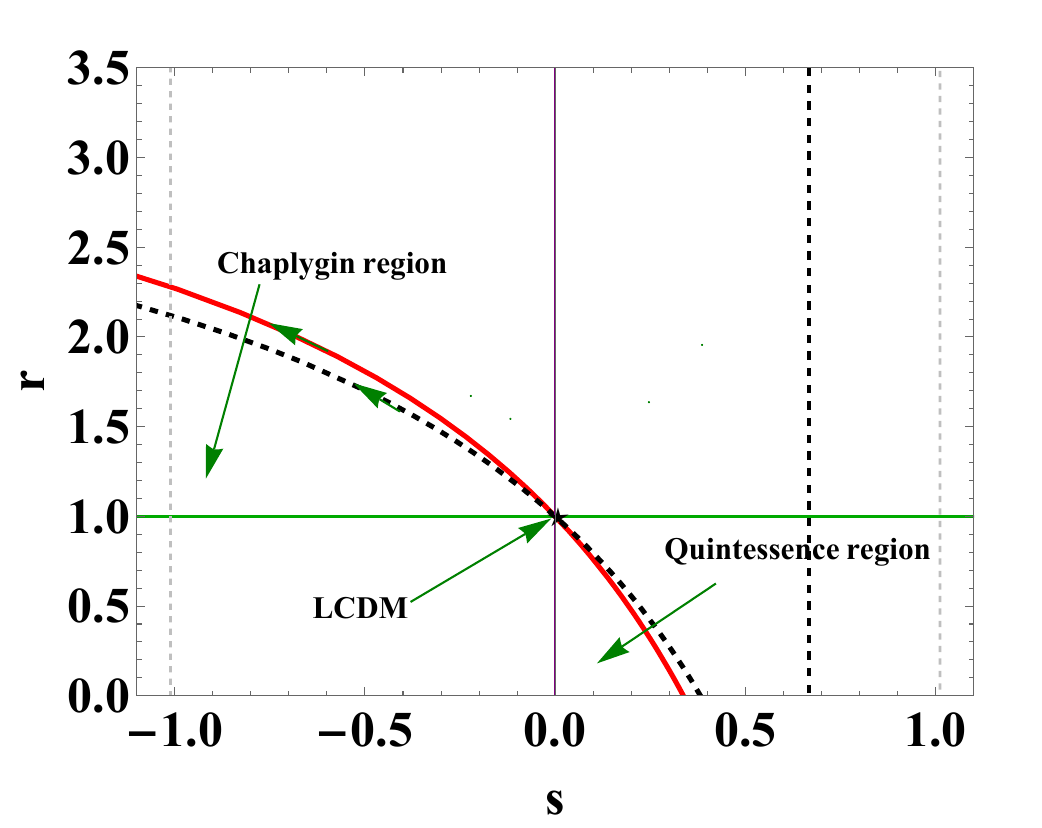}	
	(b) \includegraphics[scale=0.80]{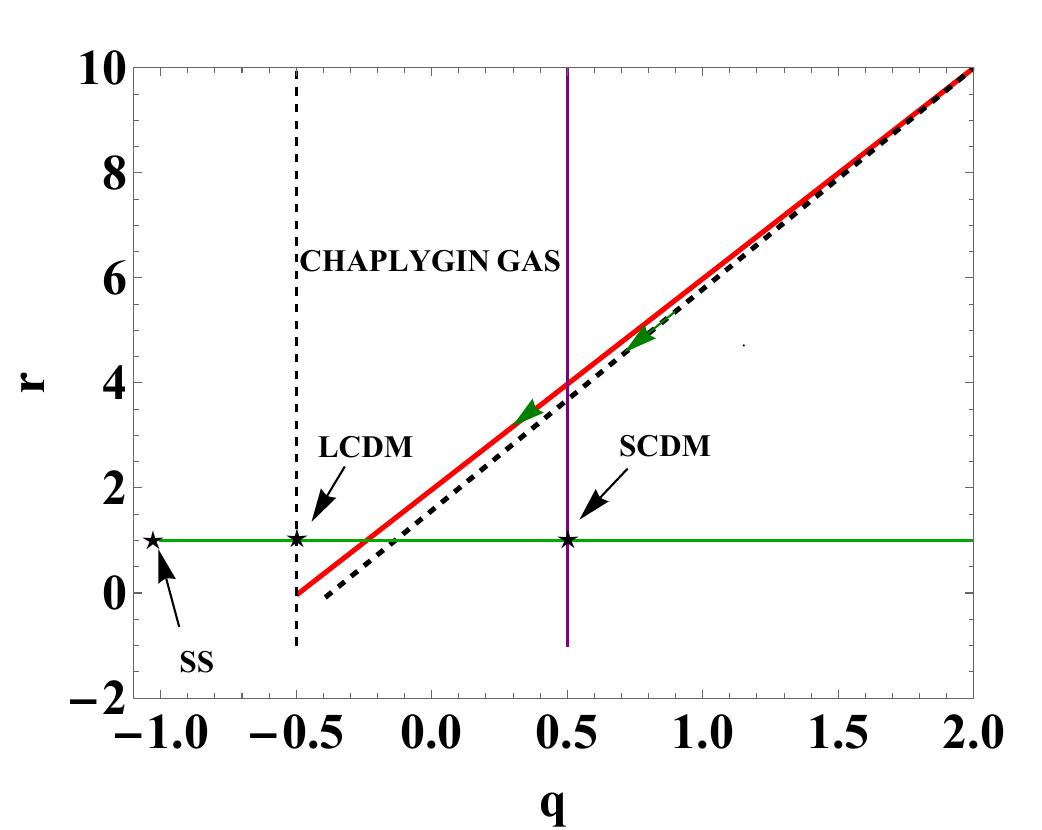}
	 \caption{(a) Plot of $ r $ vs $ s $. (b) Plot of $ r $ vs $ q $. }\label{fig10}	
\end{figure}
%%%%%%%%%%%%%%%%%%%%%%%%%%%%%%%%%%%%%%%%%%%%%%%%%%%%%%%%%%%%%%%%%%%%%%%%%%%%%%%%%%%%%%

%%%%%%%%%%%%%%%%%%%%%%%% Section 7 %%%%%%%%%%%%%%%%%%%%%%%%%%%%%%%%%%%%%
\section{Cosmography}
We examine the higher order temporal derivative of the scale factor $a$, often known as cosmographic parameters, to learn more about the evolution of the cosmos. According to \cite{ref109N59}, the Hubble $H$, deceleration $q$, jerk $j$, snap $s$, and lerk $l$ parameters are defined cosmologically using the first five-time derivatives of the scale factor $a$. 

\begin{equation}\label{32}
j= (1+z) \frac{dq}{dz}+q (1 + 2 q),
\end{equation}
\begin{equation}\label{33}
s= -(1+z) \frac{dj}{dz}-j (2 +3 q),
\end{equation}
\begin{equation}\label{34}
l= - (1+z) \frac{ds}{dz}- s (3+ 4 q).
\end{equation}
 Because it depends on the third-order derivative of the scale factor with regard to time, the jerk parameter is a highly kinematical term that more precisely characterizes the universe's expansion rate than the Hubble parameter. The universe accelerates with a positive jerk parameter \cite{ref111N60,ref112N61}.
  Blandford et al. \cite{ref113N62} explored the characteristics of the jerk parameterization in 2004, providing a different method for characterizing cosmological models that are similar to the $\Lambda$CDM model. Several key aspects of the jerk parameter have been examined by the authors \cite{ref102N52,ref107N57} in order to distinguish between various dark energy models.
  We can observe from Fig. $10$ that the resulting models change as the jerk parameter increases. $q=-1$ for the $\Lambda$CDM dark energy model, resulting in $j=1$. In the derived model, $q$ is negative and $j$ is close to $1$. Consequently, an accelerating universe near $\Lambda$CDM is represented by the derived cosmological model. The snap and lerk factors in Figures $11$ and $12$ exhibit some discontinuity in behavior as they oscillate about zero. 
  
%%%%%%%%%%%%%%%%%%%%%%%%%%%%%%%%%%%%%% Figure 8 %%%%%%%%%%%%%%%%%%%%%%%%%%%%%%%%
 \begin{figure}
\centering
	\includegraphics[scale=0.9]{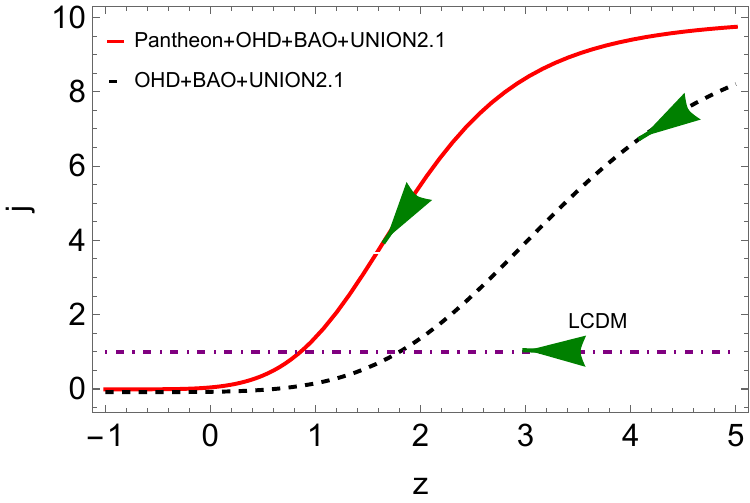}
\caption{Plot of $jerk$ versus $ z $. }\label{fig11}	
\end{figure}
 \begin{figure}
\centering
	\includegraphics[scale=0.9]{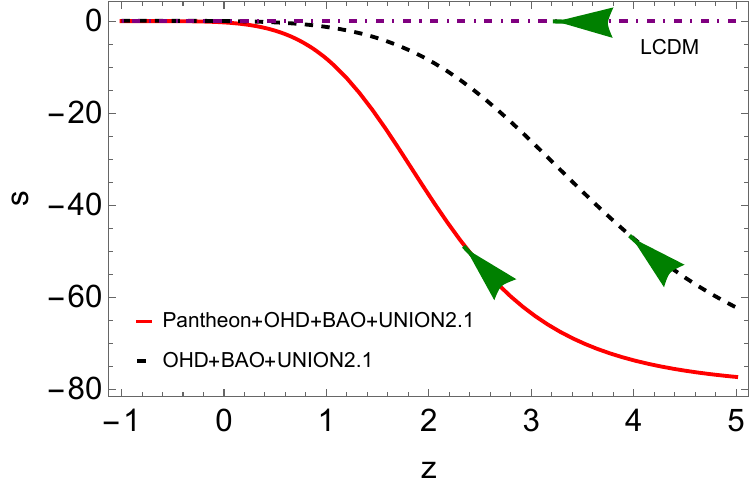}
\caption{Plot of $snap$ versus $ z $. }\label{fig12}	
\end{figure}
 \begin{figure}
\centering
	\includegraphics[scale=0.9]{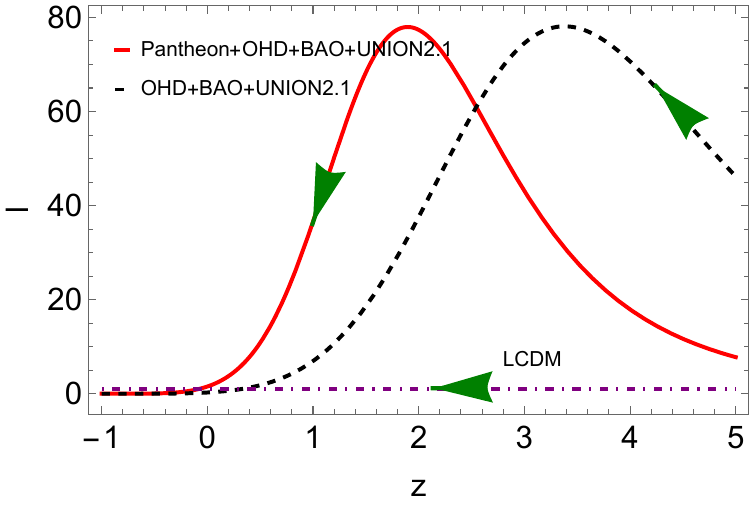}
\caption{Plot of $lerk$ versus $ z $. }\label{fig13}	
\end{figure}
%%%%%%%%%%%%%%%%%%%%%%%%%%%%%%%%%%%%%%%%%%%%%%%%%%%%%%%%%%%%%%%%%%%%%%%%%%%%%%%%%
 The physics underlying each coefficient can be inferred from the Hubble curve's shape. Specifically, the sign of $q$ determines whether the dynamics are accelerated or decelerated over the course of the universe's evolution. To put it another way, a positive acceleration parameter means that standard gravity is stronger than the other species, while a negative sign makes the standard attraction from gravity less strong. 
 Despite all of this, the dynamics cannot be fully explained by the acceleration parameter alone. The sign of $j$ is a crucial observable to ascertain whether it changes sign during its evolution. A positive jerk value, for instance, would suggest the existence of a transition moment during which the cosmos alters its expansion. The $q$ modulus tends to zero and then changes its sign in accordance with this transition. 
The two terms, $q$ and $j$, establish the local dynamics, but they are insufficient to distinguish between a cosmological model with a pure cosmological constant and one that admits an evolving dark energy element \cite{ref113aN63}. The value of $s$ is absolutely required to ascertain whether dark energy evolves at all. The functional dependency of dark energy on the redshift z is impacted by deviations from the anticipated value of $s$, as assessed in the concordance model, suggesting that it changes as the universe expands. Specifically, the terms $s$ and $l$ both affect higher orders of the Taylor expansion at greater redshifts. 
Based on these assumptions, two cosmographic regimes can be handled: the higher valid for $z \geq 2$ and the lower valid up to a redshift $z \leq 2$. For instance, the shape of the Hubble curve can be ascertained without the full list of cosmographic coefficients if $z \leq 1$. In fact, merely placing constraints on $H$, $q$, and $j$ while ignoring the other variables is sufficient to fully correct the cosmographic series. Up to $z \simeq 2$, where one even has to fix $s$, $l$, this coarse-grained method is ineffective for $z \geq 1$.
\\  
 
 Therefore, one can concentrate on the broadening of cosmographic coefficients based on the specific data sets used in the statistical study. For instance, the fluctuation of $s$ shows how the form becomes smooth at redshifts $z \geq 1$ and is mostly caused by the sign of the lerk parameter. The current data is insufficient to ensure overwhelming limits on those terms, despite the fact that a precise physical interpretation of such coefficients has been developed above \cite{ref113bN64}.
  The specific collection of data used for the research fundamentally limits the use of high-precision treatments. Usually, when one looks at a single set of cosmic data, systematics are obtained that make it impossible to place precise bounds on the cosmographic series. To obtain a cosmographic series with reduced systematic errors, a combination analysis is necessary. For instance, data in the region $z \leq 1.414$ is constrained when using the union 2.1 compilation of currently known supernova measurements. Because their meaning would only become evident at $z \geq 2$, it seems useless to constrain $s$, $l$, and $m$ with this data, or to expect to find accurate bounds on them. However, given $z > 2$ \cite{ref113cN65}, one may argue that those coefficients are more significant by referencing the differential age-independent $H$ measurements. In order to acquire the correct cosmographic series in any order, it becomes necessary to select one data set over another. Combining cosmic data to obtain tighter constraints using both low and high redshift regimes is generally a workable strategy.

%%%%%%%%%%%%%%%%%%%%%%%%%%%%Section 8 %%%%%%%%%%%%%%%%%%%%%%%%%%%%%%%%%%

\section{Om Diagnostic}
Sahni et al. \cite{ref114N66} suggest the $O_{m}$ diagnostic as an adjunct to the $\{r, s\}$ diagnostic. The geometrical diagnostic known as the Om diagnostic is specifically dependent on the Hubble parameter and redshift. The higher-order derivative of $a(t)$ is used in the study of the statefinder parameter, $(r, s)$. Since the first-order derivative only includes the Hubble parameter, it is utilized in $O_{m}$ diagnostic analysis. One simpler diagnostic is the $O_{m}$ diagnostic \cite{ref112N61}. It should be noted that the $O_{m}$ diagnostic has also been used to differentiate between various dark energy models without incorporating the EoS parameter and the density parameter of matter \cite{ref114N66,ref115N67}. This parameter configuration can be expressed as follows:

\begin{equation}
\label{31}
Om(z)= \frac{\frac{H^{2}(z)}{H_{0}}-1}{(1+z)^{3}-1}
\end{equation}

%%%%%%%%%%%%%%%%%%%%%%%%%%%%%%%%%%%%%%%%%%%%%%%%%%%%%%%%%%%%%%%%%%%%
\begin{figure}[H]
	\centering
		\includegraphics[scale=0.9]{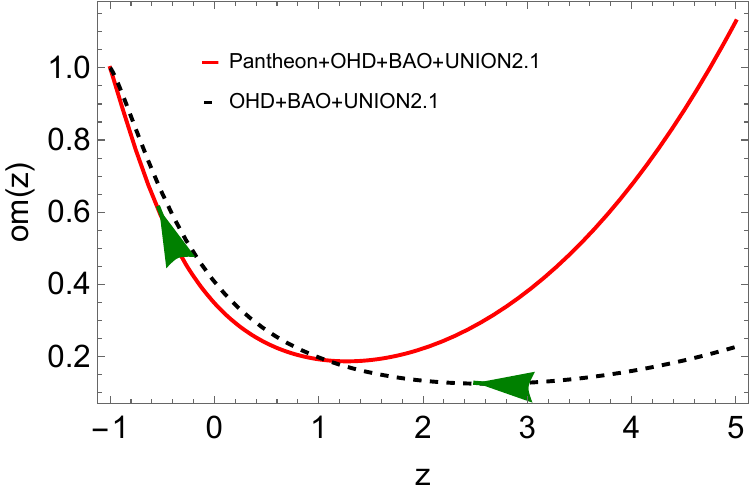}
	\caption{Evaluation of $ Om(z) $ versus $ z $. }\label{fig15}
\end{figure}\l{fig14}
%%%%%%%%%%%%%%%%%%%%%%%%%%%%%%%%%%%%%%%%%%%%%%%%%%%%%%%%%%%%%%%
The $O_{m}(z)$ evolution with redshift is plotted in figure $13$.
While the negative curve suggests that DE behaves like quintessence $(\omega >-1)$, the positive curve of the $O_{m}(z)$ trajectories displays the phantom behavior $(\omega <-1)$. For a phantom EoS, we discover that $O_{m}(z)$ has a positive slope (curvature), which is a general characteristic of the dark energy model with $\omega < -1$. For any value of the matter density, the phantom's positive curvature separates the cosmological from zero curvature $\Lambda$CDM model (Fig. $13$).

%%%%%%%%%%%%%%%%%%%%%%%%%%%%%%%%%%%%%%%%%%%%%%%%%%%%%%%%%%%%%%%

\section{Analysis of the $\omega-\omega^{'}$ pair}
%%%%%%%%%%%%%%%%%%%%%%%%%%%%%%%%%%%%%%%%%%%%%%%%%%%%%%%%%%%%%
\begin{figure}	
	\begin{center}
				\includegraphics[width=15cm,height=9cm, angle=0]{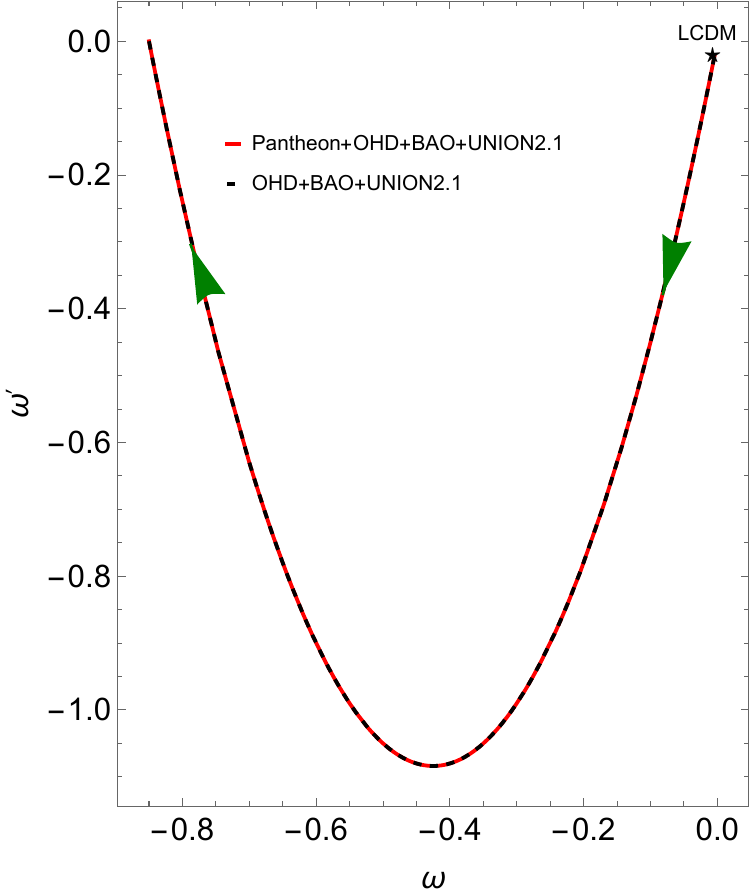}
		\caption {  Analysis of $\omega-\omega'$ plane.} 
	\end{center}
\end{figure}\l{fig15}
%%%%%%%%%%%%%%%%%%%%%%%%%%%%%%%%%%%%%%%%%%%%%%%%%%%%%%%%%%%%%%%

Lastly, we looked into alternative dynamical diagnosis for our model, which is also frequently used in the literature, $\omega- \omega^{'}$, where prime displays the derivative with regard to $ \log a$. In cosmological model analysis, the $\omega$ - ${\omega}^{\prime}$ pair typically refers to the equation of state (EoS) parameter of dark energy and its evolution. $\omega$ is the equation of state parameter of dark energy, defined as the ratio of its pressure $p$ to its energy density $\rho$, i.e., $\omega = \frac{p}{\rho}$. In the standard $\Lambda$CDM model, dark energy is modeled as a cosmological constant with $\omega = -1$. However, in dynamical dark energy models, $\omega$ can vary with time. ${\omega}^{\prime}$ (or $\frac{d\omega}{da}$) represents the rate of change of the equation of state parameter with respect to the scale factor $a$ (or sometimes redshift $z$). It quantifies the evolution of dark energy and helps distinguish between different dark energy models. The $\omega$ - ${\omega}^{\prime}$ parameter space is useful for testing and constraining different theoretical models of dark energy, such as quintessence, k-essence, or modified gravity theories. In the $\omega-\omega^{'}$ figure, the typical $\Lambda$CDM is indicated by the fixed points $\omega = -1$, $\omega^{'} = 0$. In the $\omega-\omega^{'}$ plane, the extent of the DE quintessence model has been examined in \cite{ref4aN68}. The $\omega-\omega^{'}$ plane has been used to study several DE models \cite{ref5aN69,ref6aN70,ref7aN71}. 
Caldwell et al.\cite{ref4aN68} state that there are two different types of $\omega-{\omega}^{\prime}$ planes. By showing ${\omega}^{\prime}<0$ at $\omega<0$, the cooling model simulates the Universe's accelerated expansion, while the thawing model supports the Universe's decelerated expansion by explaining the range ${\omega}^{\prime}>0$ at $\omega<0$.
\\
 
  Figure 14 displays our model's evolutionary path over the $\omega-{\omega}^{\prime}$ plane for a range of model parameter values. This shows that our model's evolutionary trajectory begins at the $\Lambda$CDM fixed point $\omega = -1$, $\omega^{'} = 0$, and ends at $\omega<0$ in the freezing region ${\omega}^{\prime}<0$. The phantom divide line $\omega = -1$ is also crossed by $\omega$.  As a result, it exhibits chaplygin gas characteristics, and cosmic expansion is currently speeding up.  Model parameters only have quantitative effects.
  \\

\section{Conclusion:} 
Our study aims to examine the effects of the bulk viscosity  in the standard cosmic pressure on the universe's evolution phase. We explore an axially symmetric Bianchi type-I model of the universe with bulk viscous fluid as a source of gravitational field under the framework of Einstein's field equations. Viscous pressure associated with bulk viscosity can act like a negative pressure in a cosmological model. Since negative pressure causes cosmic acceleration and defies gravitational attraction, it is a characteristic of dark energy. In our model, Barotropic bulk viscous pressure is assumed to be $-3\zeta H^2$. The following is a summary of the paper's highlights:
 
\begin{itemize}
    \item We demonstrate a transitional universe that began to accelerate smoothly at a specific red shift after initially decelerating. 
    
    \item The Hubble, deceleration, and Om parameters, as well as the jerk and lurk parameters and, most importantly, the universe's density, have all been addressed. 
    
    \item  The model is made under the constraints of the four data sets. The Hubble 46 data set(OHD) describes Hubble parameter values at various redshifts . Union 2.1 compilation data sets (Union) comprising of distance modulus of 580 SNIa supernovae at different redshifts. The Pantheon data set(Pan) which contains apparent magnitudes of 1048 SNIa supernovae at various redshifts and finally BAO data set of volume averaged distances at 5 redshifts. These data sets and their combinations were used to estimate the model parameters $H_0$, $l$ and $\zeta.$
    The OHD+BAO~and~OHD+Pan+BAO+Union combined data sets provide the best fit Hubble parameter value $H_0$ as $66.912 ^{+0.497}_{-0.501})$ Km/s/Mpc and $74.216 ^{+0.150}_{-0.148}$ Km/s/Mpc respectively. 
    
    \item We have performed state finder diagnostics to discuss the nature of dark energy. Some other geometrical parameters like the Jerk parameter and the Om diagnostic are also being discussed to clarify the nature of the dark energy model. 
    The study reveals that the model behaves like a quintessence in late time and approaches the $\Lambda$ CDM model.
    \end{itemize}
In summary, the bulk viscosity is a crucial concept in exploring the deeper nature of dark energy and how it governs the expansion and ultimate fate of the universe. It offers a more flexible framework than a static $\Lambda$. We should now talk about future observational experiments that could further constrain the model. Several existing and forthcoming observational efforts aim to strengthen cosmological parameter constraints, further our understanding of dark energy, dark matter, and cosmic expansion, and test alternative gravity theories. Large-Scale Structure (LSS) and Galaxy Surveys - DESI (Dark Energy Spectroscopic Instrument, 2021-Present). Refining dark energy constraints (e.g., Euclid, DESI, Roman, CMB-S4). Solving the Hubble tension (e.g., Roman, LISA, gravitational wave standard sirens).

\section*{Declaration of competing interest}
The authors declare that they have no known competing financial interests or personal relationships that could have appeared to influence the work reported in this paper.

\section*{Data availability}
No data was used for the research described in the article.

\section*{Acknowledgments}
The author, A. Pradhan, thanks the IUCAA in Pune, India, for providing the facility through the Visiting Associateship programs. The authors thank the reviewers for their valuable comments and recommendations, which improved the manuscript in its current form.

\end{document}